\documentclass[conference]{IEEEtran}

\ifCLASSINFOpdf
\usepackage[pdftex]{graphicx}
\else
\fi

\usepackage{amssymb}
\usepackage{amsmath}

\usepackage{algorithm}%
\usepackage[noend]{algpseudocode}
\usepackage{algorithmicx}

\ifCLASSOPTIONcompsoc
  \usepackage[caption=false,font=normalsize,labelfont=sf,textfont=sf]{subfig}
\else
  \usepackage[caption=false,font=footnotesize]{subfig}
\fi

\usepackage[usenames,dvipsnames,table]{xcolor}
\usepackage{ifthen}

\usepackage{booktabs}

\usepackage{acronym}

\usepackage{tcolorbox}
\definecolor{sgx}{HTML}{F4CCCC}
\tcbset{
    title=Enclaved,
    arc=0mm,
    boxsep=0.6mm,
    top=0.6mm,
    bottom=0.2mm,
    left=5mm,
    width=\linewidth+5.6mm,
    enlarge left by=-5.8mm,
    colframe=sgx,
    coltitle=black,
    boxrule=0.2mm,
    halign title=right,
}

\usepackage[style=ieee,backend=bibtex,natbib=true,sorting=none,doi=true,isbn=false,alldates=year]{biblatex}
\bibliography{references}

\makeatletter
\renewcommand{\ALG@beginalgorithmic}{\small}
\makeatother

\usepackage{tikz}
\usepackage{hyperref}
\newcommand\copyrighttext{%
  \footnotesize \textcopyright 2018 IEEE.
    Personal use of this material is permitted.
    Permission from IEEE must be obtained for all other uses,
    in any current or future media, including reprinting/republishing this
    material for advertising or promotional purposes, creating new collective
    works, for resale or redistribution to servers or
    lists, or reuse of any copyrighted component of this work in other works.
    Pre-print version. Presented in the \href{https://dsn2018.uni.lu/}{48th IEEE/IFIP International Conference on Dependable Systems and Networks (DSN '18)}. For the final published version, refer to DOI \href{https://doi.org/10.1109/DSN.2018.00032}{10.1109/DSN.2018.00032}}
\newcommand\copyrightnotice{%
\begin{tikzpicture}[remember picture,overlay]
\node[anchor=south,yshift=10pt,fill=yellow!20] at (current page.south) {\fbox{\parbox{\dimexpr\textwidth-\fboxsep-\fboxrule\relax}{\copyrighttext}}};
\end{tikzpicture}%
}
\begin{document}
\title{IBBE-SGX: Cryptographic Group Access Control using Trusted Execution Environments}

\author{\IEEEauthorblockN{Stefan Contiu\IEEEauthorrefmark{1}\IEEEauthorrefmark{3},
Rafael Pires\IEEEauthorrefmark{2},
S\'ebastien Vaucher\IEEEauthorrefmark{2}, 
Marcelo Pasin\IEEEauthorrefmark{2}, 
Pascal Felber\IEEEauthorrefmark{2} and 
Laurent R\'eveill\`ere\IEEEauthorrefmark{1}}
\vspace{1mm}
\IEEEauthorblockA{
	\IEEEauthorrefmark{1}University of Bordeaux, France, \texttt{firstname.lastname@u-bordeaux.fr}\\
	\IEEEauthorrefmark{2}University of Neuch\^atel, Switzerland, \texttt{firstname.lastname@unine.ch}\\
	\IEEEauthorrefmark{3}Scille SAS, France
}
\vspace{-0.9cm}
}

\maketitle
\copyrightnotice

\pagestyle{plain}

\acrodef{IBBE}{Identity-Based Broadcasting Encryption}
\acrodefindefinite{IBBE}{an}{an}
\acrodef{HE}{Hybrid Encryption}
\acrodefindefinite{HE}{an}{an}
\acrodef{CA}{Certificate Authority}
\acrodefplural{CA}{Certificate Authorities}
\acrodef{TCB}{Trusted Computing Base}
\acrodef{IAS}{Intel Attestation Service}
\acrodef{SGX}{Software Guard Extensions}
\acrodef{EPC}{Enclave Page Cache}
\acrodef{TLS}{Transport Layer Security}
\acrodef{PKI}{Public Key Infrastructure}
\acrodef{TEE}{Trusted Execution Environment}
\acrodef{AES}{Advanced Encryption Standard}
\acrodef{ECC}{Elliptic Curve Cryptography}
\acrodef{HE-PKI}{Hybrid Encryption with Public Key}
\acrodef{HE-IBE}{Hybrid Encryption with Identity-based Encryption}
\acrodef{BE}{Broadcast Encryption}
\acrodef{TA}{Trusted Authority}
\acrodefplural{TA}{Trusted Authorities}
\acrodef{ABE}{Attribute-Based Encryption}
\acrodef{HIBE}{Hierarchical Identity Based Encryption}
\acrodef{FE}{Functional Encryption}
\acrodef{IBE}{Identity Based Encryption}
\acrodef{API}{Application Programming Interface}
\acrodef{LoC}{Line of Code}
\acrodefplural{LoC}{Lines of Code}
\acrodef{CDF}{Cumulative Density Function}

\begin{abstract}

While many cloud storage systems allow users to protect their data by making use of encryption, only few support collaborative editing on that data.
A major challenge for enabling such collaboration is the need to enforce cryptographic access control policies in a secure and efficient manner.
In this paper, we introduce IBBE-SGX, a new cryptographic access control extension that is efficient both in terms of computation and storage even when processing large and dynamic workloads of membership operations, while at the same time offering \emph{zero knowledge} guarantees.

IBBE-SGX builds upon \ac{IBBE}.
We address \ac{IBBE}'s impracticality for cloud deployments by exploiting Intel \ac{SGX} to derive cuts in the computational complexity.
Moreover, we propose a group partitioning mechanism such that the computational cost of membership update is bound to a fixed constant partition size rather than the size of the whole group.
We have implemented and evaluated our new access control extension.
Results highlight that IBBE-SGX performs membership changes 1.2 orders of magnitude faster than the traditional approach of \ac{HE}, producing group metadata that are 6 orders of magnitude smaller than \ac{HE}, while at the same time offering \emph{zero knowledge} guarantees.

\end{abstract}

\acresetall

\IEEEpeerreviewmaketitle

\section{Introduction}
\label{sec:intro}

Cloud storage services such as Amazon Web Services, Google Cloud Platform or Microsoft Azure have shown rapid adoption during the last years~\cite{seybert_internet_2014}.
However, they lack in offering trustworthiness and confidentiality guarantees to end users.
Although threats commonly originate from malicious adversaries breaching security measures to gain access to user data,
the menace can also come from an employee with generous privileges,
or curious governments warranting data collection for national interest.
To overcome these issues, many approaches rely on the construction of cryptographic solutions in which the data is secured on the client side before reaching the storage premises~\cite{bessani2014scfs}, therefore mitigating the lack of trust in the cloud provider.

To enable collaborative operations on the already secured data, one needs to enforce access control policies. 
Because of the untrusted nature of cloud storage, such administrative operations also need to be cryptographically protected. 
Enforcing cryptographic access control on an untrusted cloud storage context is subject to a number of requirements.
First, access control schemes should incur as low traffic overhead as possible because cloud storages have slower response times in comparison to traditional storage mediums.
Second, since a realistic and dynamic membership operations pattern~\cite{Garrison:2016:DACCloud} coupled with large volumes of users can make the system unusable in practice, the system must be limited to an acceptable computational bound.
Third, as only privileged users perform access control operations, it is required that they gain \textit{zero knowledge} to the data content to which the policy is applied.
Last, identities normally employed by users when interacting with the cloud storage (i.e. existing credentials) should be sufficient for membership operations, hence avoiding complex trust establishment protocols.

A number of cryptographic constructions have been proposed for achieving access control.
The simplest one, popularly referred to as \ac{HE}, makes use of symmetric and public-key cryptography, by employing the former on the actual data and the latter on the symmetric key~\cite{goh2003sirius}.
Other approaches rely on pairing-based cryptography as a substitute for public-key cryptography, and offer different levels of granularity for specifying access control policies. 
A few examples include: 
\ac{IBE}~\cite{boneh2001identity} which works similarly to public-key encryption at the identity level;
\ac{ABE}~\cite{goyal2006attribute} that supports a richer tree-like access policy expressiveness;
or \ac{IBBE}~\cite{sakai2007identity} that can capture group-like policies.
Similarly, \emph{Identity-based proxy re-encryption} relies on a semi-trusted middle entity to whom users delegate the re-encryption rights~\cite{green2007identity}.

Unfortunately, pairing-based constructions suffer from important performance issues. 
According to \textit{Garrison et al.}~\cite{Garrison:2016:DACCloud}, they are an order of magnitude slower than public-key cryptography.
Remarkably, even \ac{HE-PKI} incurs prohibitive costs for dynamic access control (see Figure~\ref{fig:bb_test}).
Moreover, the aforementioned constructions do not guarantee our \textit{zero knowledge} requirement.

In this paper, we introduce a new cryptographic access control scheme that is both computationally- and storage-efficient considering a dynamic and large set of membership operations, while offering \textit{zero knowledge} guarantees.
\textit{Zero knowledge} is guaranteed by executing the cryptographic access control membership operations in \iac{TEE}.\footnote{Only membership operations rely on the \ac{TEE}; user operations are done in a conventional execution environment.}
Our scheme is based on \ac{IBBE} which is known to be flexible enough to produce small constant policy sizes.
Its main drawback is its high impracticable computational cost.
Our solution is to execute the membership operations of \ac{IBBE} within the \ac{TEE} so that we can make use of a master secret key.
The \ac{TEE} guarantees that this secret stays within the trusted computing boundary.
We can therefore propose an optimization of a well-studied \ac{IBBE} scheme~\cite{delerablee2007identity} that drastically reduces its computational complexity.
A remaining issue is the computational complexity required for users to derive membership changes.
To mitigate this last aspect, we propose a group partitioning mechanism such that the computational cost on the user-side is bound to a fixed constant partition size rather than the potential large group size.

We have implemented our new access control scheme using Intel \ac{SGX} as \ac{TEE}.
To the best of our knowledge, we are the first to adapt the pairing-based-specific library PBC~\cite{lynn2006pbc} and its underlying dependency GMP~\cite{granlund1991gmp} to accurately run within \ac{SGX}.
Moreover, we deployed our system on a commercially-available public cloud storage.
Our evaluation shows that our scheme only requires few resources while performing better than \ac{HE}, in addition to providing \textit{zero knowledge}. 

To summarize, our contributions are the following:

\begin{itemize}

\item 
We propose a new approach to \ac{IBBE} by confiding in Intel \ac{SGX}.
To the best of our knowledge, this is the first effort seeking to lower the computational complexity bound of a well-studied \ac{IBBE} scheme using \acp{TEE} as enabling technology.
Additionally, our scheme only requires \ac{TEE} support for a minimal set of users (i.e. administrators).

\item 
We instantiate the novel IBBE-SGX construction to an access control system hosted on a honest-but-curious cloud storage, proposing an original partitioning scheme that lowers the time required by users to ingest access control changes.

\item We fully implemented our original access control system and evaluated it against a realistic setup.
We conducted extensive evaluations, showing that our system surpasses the performances obtained using state-of-the-art approaches.

\end{itemize}

Even though the main motivation for this work is to securely share data in a cloud environment, the proposed solution can be applied for encrypting arbitrary information that is securely broadcasted to a group of users over any shared media.
Some other examples, besides cloud storages, are peer-to-peer networks or pay-per-view TV. 

The rest of the paper is organized as follows.
Section~\ref{sec:model} presents the model assumptions undertaken within the problem context. 
We provide some background about Intel \ac{SGX} and cryptographic schemes in Section~\ref{sec:building_blocks}.
Section~\ref{sec:solution} presents the unique constructions that allow us to lower the complexity of an access control scheme by relying on \iac{TEE}.
In the second part of that section, details about the partitioning mechanism are shown.
We describe the design and implementation of an end-to-end system built on top of the scheme in section~\ref{sec:implementation}.
Section~\ref{sec:evaluation} presents the evaluation of our solution by performing both micro- and macro-benchmarks.
Section~\ref{sec:related_work} presents the related work in the fields of cryptography, access control systems and \ac{SGX}.
Finally, Section~\ref{sec:conclusion} concludes and presents future work.
\section{Model}
\label{sec:model}

\begin{figure}[t]
	\centering
	\includegraphics[width=0.4\textwidth]{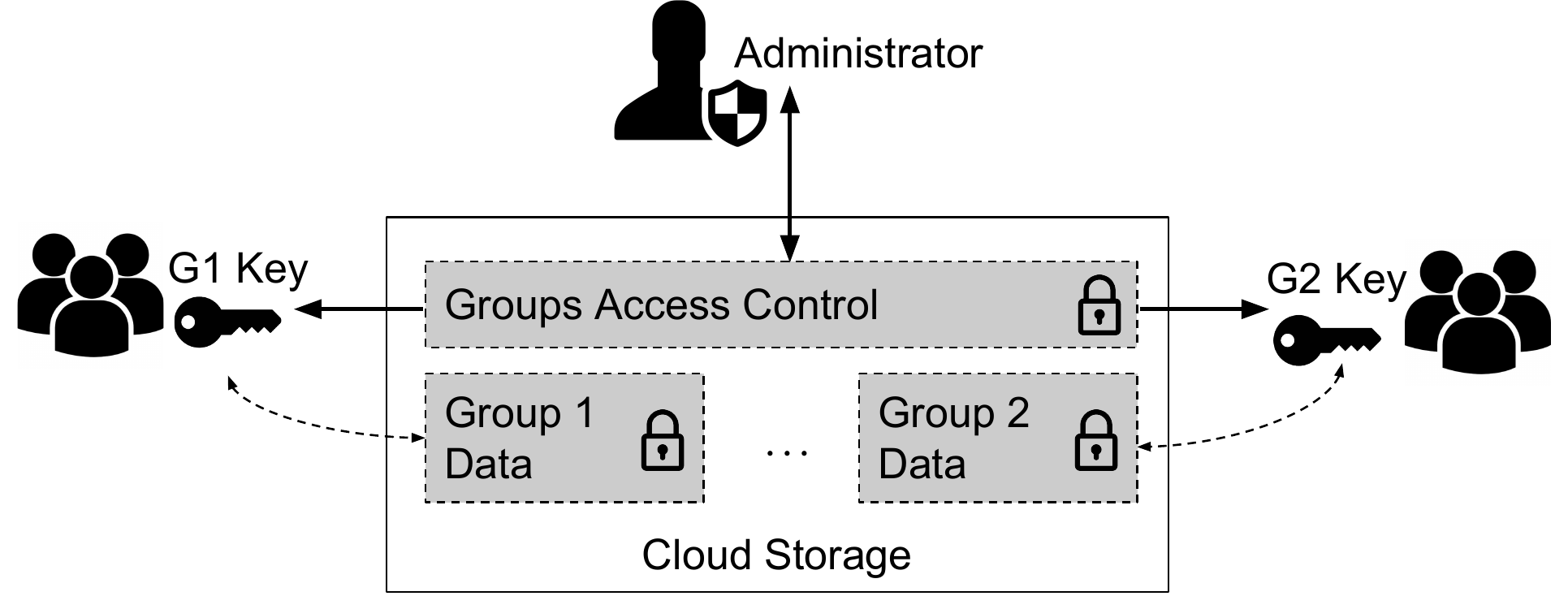}
	\caption{\label{fig:model_diagram}Model diagram.}
	\vspace{-0.6cm}
\end{figure}

In this study, we consider groups of users who perform collaborative editing on cryptographically-protected data stored on untrusted cloud storage systems.
The data is protected using a block cipher encryption algorithm such as \ac{AES} making use of a symmetric group key \texttt{gk}.
As illustrated in Figure~\ref{fig:model_diagram}, this work addresses the  challenge of designing a system for group access control, in which the group key \texttt{gk} is cryptographically protected and derivable only by the members of the group.
Because groups may become large with a significant turnover in their members, we investigate the implication of numerous member additions and revocations happening throughout their lifetimes.

We distinguish between two types of actors interacting within the system: \textit{administrators} and \textit{users}.
All group membership operations are performed by administrators.
Their duties include creating groups, and adding or revoking group members.
The administrators manifest an honest-but-curious behavior, correctly serving work requests but with a possible malicious intent of discovering the group key \texttt{gk}.
On the other hand, users listen to the cloud storage for group membership changes, and derive the new group key \texttt{gk} whenever it changes. 
Users are considered of having a trusted behavior.

The role of the cloud storage is to store the definitions of groups access control, also referred to as groups metadata, together with the list of members composing the group, and the actual group data. 
Besides being a storage medium, we also use the cloud storage as a broadcasting interface for group access control changes.
Administrators are communicating with the cloud each time a group membership operation takes place so that users can be notified of the group membership update.
We consider the cloud storage to show a similar behavior than administrators (i.e. honest-but-curious).
It correctly services assigned tasks albeit with a possible malicious intent of peeking into the groups secrets.
Moreover, when manifesting the curious behavior, the cloud storage could collude with any number of curious administrators or revoked users.

Size-wise, we target a solution that can accommodate groups of a very large-scale nature.\footnote{Our evaluation uses 1 million users as the largest group size.}
It is desired that administrators perform membership changes for multiple groups at a time, therefore the number of administrators is relatively small when compared to the number of users.

We choose to ensure authenticity guarantees only with respect to administrator identities, and therefore authenticate membership changes operations.
Also, authenticating the group data created by users is out of scope, the current model being focused on confidentiality guarantees.
Therefore, the notion of a reference monitor~\cite{sandhu1994access, Garrison:2016:DACCloud} on the cloud storage is not pertinent within our context.
Finally, we do not consider hiding the identities of group members, nor the type of executed membership operations, as they can be inferred by the cloud storage from traffic access patterns.
Privacy constructions offering such guarantees~\cite{mayberry2014efficient,devadas2016onion,apon2014verifiable} are orthogonal to our work.

\section{Background}
\label{sec:building_blocks}
The building blocks that lead to the creation of our access control extension are Intel SGX, Hybrid Encryption, and Identity Based Broadcast Encryption. 
This background section presents them together with open challenges.

\subsection{Intel SGX}
\label{subsec:backgroundsgx}

Intel \ac{SGX} is an instruction extension available on modern x86 CPUs manufactured by Intel.
Similarly to ARM Trustzone~\cite{alves2004trust} or Sanctum~\cite{costan2016sanctum}, \ac{SGX} aims to shield code execution against attacks from privileged code (e.g. infected operating system) and certain physical attacks.
A unit of code protected by \ac{SGX} is called an \emph{enclave}.
Computations done inside the enclave cannot be seen from the outside \cite{costan2016intel}.
\ac{SGX} seamlessly encrypts memory so that plaintext data is only present inside the CPU package.
The assumption is that opening the CPU package is difficult for an attacker, and leaves clear evidence of the breach.
Encrypted memory is provided in a processor-reserved memory area called the \ac{EPC}, which is limited to 128\,MB in the current version of \ac{SGX}.

Intel provides a way for enclaves to \emph{attest} each other~\cite{sgx-sdk}.
After the attestation process, enclaves will be sure that each other is running the code that they are meant to execute.
The attestation process can be extended to \emph{remote attestation} that allows a piece of software running on a different machine to make sure that a given enclave is running on a genuine Intel SGX-capable CPU.
An Intel-provided online service --- the \ac{IAS} --- is used to check the signature affixed to a \emph{quote} created by the CPU~\cite{sgx-sdk}.
As part of the attestation process, it is possible to provision the enclave with secrets.
They will be securely transmitted to the enclave if and only if the remote attestation process succeeds.

The \ac{TCB} of an SGX enclave is composed of the CPU itself, and the code running within.
The assumption is that we trust Intel for securely implementing \ac{SGX}.
Nevertheless, it has been shown that \ac{SGX} is vulnerable to side-channel attacks~\cite{206170}.
We consider this flaw to be orthogonal to our research, and hence do not consider it in our security evaluation.

\subsection{A Naive Approach to Group Access Control}

Suppose that we want to come up with a simple, yet secure, cryptographic scheme to protect a group key \texttt{gk}.
We can make use of an asymmetrical encryption primitive~\cite{ferguson2003practical}, based on RSA or \ac{ECC}.
As each user in the system possesses a public-private key pair, the scheme consists in encrypting \texttt{gk} using the public key of each member in the group.
A member of the group can then deduce \texttt{gk} by decrypting the resulting ciphertext using her private key.
This construction is sometimes referred to as \acf{HE}~\cite{Garrison:2016:DACCloud}, or Trivial Broadcast Encryption Scheme~\cite{stinson2005cryptography}.

To achieve the \textit{zero knowledge} requirement, administrators could be asked to run \ac{HE} within an \ac{SGX} enclave, thus protecting the discovery of \texttt{gk}.
However, before discussing the cost of such an integration between \ac{HE} and \ac{SGX}, we point out a number of prior weaknesses of \ac{HE}.

First, the amount of group metadata grows linearly with the number of members in the group, making it impractical in the context of very large groups.
Second, when revoking group members, a new key \texttt{gk} needs to be created; the entire group metadata also needs to be generated again by encrypting the latest value of \texttt{gk}.
As the group size increases, the computational cost of the scheme grows linearly.
Likewise, the latency incurred for putting, getting and storing the group metadata on the cloud storage will also seriously expand.

Furthermore, when performing group membership operations, the administrators need to entrust the authenticity of the public keys linked to the identity of the members.
\acp{PKI}~\cite{ferguson2003practical} can be used to solve this issue.
Besides the trust risks that the \ac{PKI} brings~\cite{ellison2000ten}, one needs to account for the practical costs of setting up, running and accessing \iac{PKI}.
To mitigate these risks, one could choose to substitute public-key primitives with identity-based ones.
\acf{IBE}~\cite{boneh2001identity, waters2005efficient} makes use of arbitrary strings as public keys; we can therefore use a user name directly as a public key.
The user secret key is generated at setup phase or later by \iac{TA}.
Obviously, both \ac{HE-PKI} and \ac{HE-IBE} have the same inner functioning, when making abstraction of the key methodology choice.

Integrating \ac{SGX} with \ac{HE-PKI} and \ac{HE-IBE} is required in order to guarantee the \textit{zero knowledge} property against administrators.
As hybrid encryption is causing a high group metadata expansion, it has a direct impact on the memory space that is used inside the SGX enclave.
Accessing memory in \ac{SGX} enclaves can induce an overhead of up to 19.5\,\% for write accesses and up to 102\,\% for read accesses~\cite{weisse2017regaining}. 
Apprehensive about the hypothesized \ac{SGX} degradation in performance caused by the group metadata expansion, we shift the focus on finding a solution with minimal expansion.

\subsection{\acf{BE} and Identity-Based \acs{BE}}

In order to optimize both \ac{SGX} and cloud transit costs, we investigate the possibility of cryptographic schemes that  induce a minimal group expansion.

\acf{BE}~\cite{fiat1993broadcast} is a public-key cryptosystem with a unique public key that envelopes the entire system, contrary to the \ac{HE} scheme where each user uses a different public key.
However, each user in a \ac{BE} system has a unique private key generated by a trusted authority.
To randomly generate a group key \texttt{gk} and the associated group metadata (named \textit{encrypt} operation within \ac{BE} systems), one makes use of the system-wide public key.
On the other side, when a user wants to reveal \texttt{gk} (\textit{decrypt} in \ac{BE} systems), she makes use of her individual private key.

As broadcast encryption schemes come with different contextual models, we impose a number of conditions.
First, to maintain the same threat model as \ac{HE}, we are only investigating the use of fully collusion-resistant \ac{BE} schemes~\cite{boneh2005collusion}, in which no coalition of members outside of the group could reveal \texttt{gk}.
Second, the set of users participating in the system is not initially known, thus we rely on the usage of \textit{dynamic} \ac{BE} schemes~\cite{delerablee2007fully}.
Third, as in the case of \ac{HE}, we would prefer constructions that can accommodate the use of \ac{IBE}.

Piercing through the existing research literature, we identified \iac{IBBE} scheme~\cite{Delerablee:2007:IBBE} that not only fulfills all the aforementioned requirements, but also operates with group metadata expansions and user private keys of constant sizes.
Moreover, the scheme has an additional strategic advantage that proves beneficial in our context: the system-wide public key size is linear in the maximal size of any group.

Upon analyzing the computational complexity of the selected \ac{IBBE} scheme, one can notice that creating \texttt{gk} given a set of members, as well as decrypting it as a user, are operations with a quadratic complexity in the number of members.
Therefore, even though the scheme brings a tremendous gain in the size of group metadata expansion, the computational cost of \ac{IBBE} might be excessive for practical use.

\begin{figure}[!t]
	\centering
	\subfloat{\includegraphics[width=0.7\columnwidth]{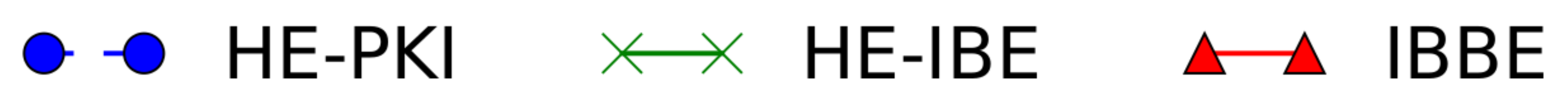}}
	\\[-2ex]
	\addtocounter{subfigure}{-1}
	\subfloat[Latency for group creation.]{\includegraphics[width=0.5\linewidth]{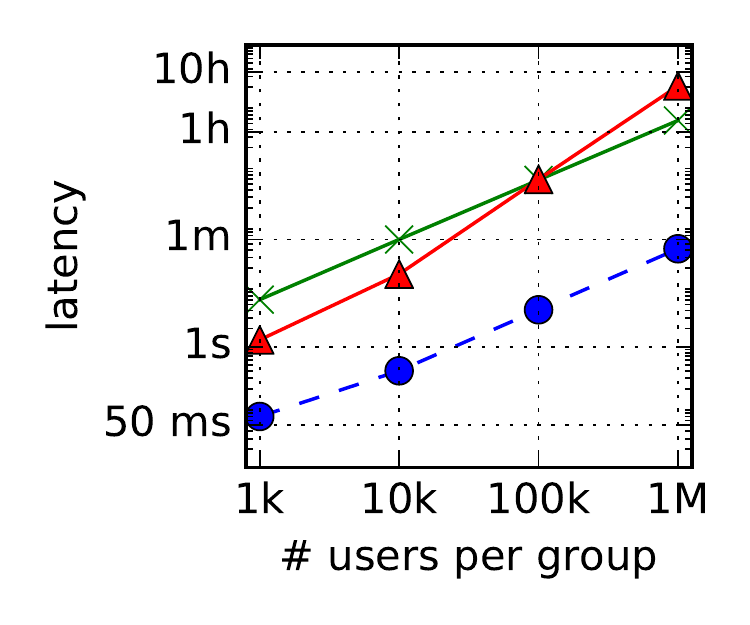}%
		\label{fig:bb_lat}}
	\hfil
	\subfloat[Group metadata expansion.]{\includegraphics[width=0.5\linewidth]{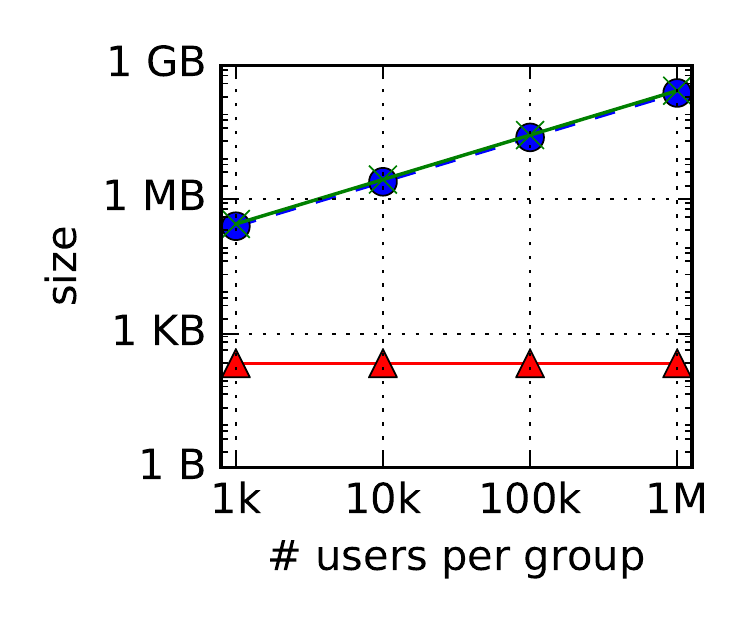}%
		\label{fig:bb_size}}
	\caption{Performance of \ac{HE-PKI}, \ac{HE-IBE} and \ac{IBBE}, without \textit{zero knowledge}.}
	\label{fig:bb_test}
\end{figure}

Figure \ref{fig:bb_test} exemplifies the performance of \ac{HE-PKI}, \ac{HE-IBE} and \ac{IBBE} schemes in their raw form, before any integration with \ac{SGX} is considered.
The sub-figure on the left displays the total time taken for the operation of creating a group while the one on the right shows the size occupied by the expansion of group metadata.
The optimality of \ac{IBBE} regarding the size of group metadata expansion is immediately obvious.
It always produces 256\,bytes of metadata, regardless of the number of users per group. 
That is preferable compared to \ac{HE-PKI} and \ac{HE-IBE}, which produce 
increasingly larger values, as much as 27\,MB for groups of 100,000 users, and 274\,MB for the largest benchmarked group size.
On the other hand, \ac{IBBE} performs much worse than \ac{HE-PKI} when considering the 
execution time. It is 150$\times$ and 144$\times$ slower for groups of 10,000 and 100,000 users, respectively.

There is no doubt that running the \ac{IBBE} scheme in this form is inadequate.
In the remainder of this paper, we describe two original contributions, one that changes a traditional assumption of the \ac{IBBE} scheme, and a second that lowers the user decryption time.

\section{IBBE-SGX}
\label{sec:solution}

IBBE-SGX can be broadly described in 3 steps: 
(i)~trust establishment and private key provisioning; 
(ii)~membership definitions and group key provisioning;
and (iii)~membership changes and key updates.

\subsection{Trust Establishment}

\ac{IBBE} schemes generate a single public key that can be paired with several 
private keys, one per user. 
Users, in turn, need to be sure that the private key they receive is indeed 
generated by someone they trust, otherwise they would be vulnerable to malicious
entities trying to impersonate the key issuer.
To achieve that, we rely upon \iac{PKI} to provide
verifiable private keys to users.

Another security requirement of IBBE-SGX is that the key management must be
kept in \iac{TEE}.
Therefore, there must be a way of checking whether that is the case.
On that front, Intel \ac{SGX} makes it possible to attest enclaves.
Running this procedure gives the assurance that a given piece of binary code
is truly the one running within an enclave, on a genuine Intel \ac{SGX}-capable processor
(Section~\ref{subsec:backgroundsgx}).

\begin{figure}
    \centering
    \includegraphics[width=0.5\textwidth]{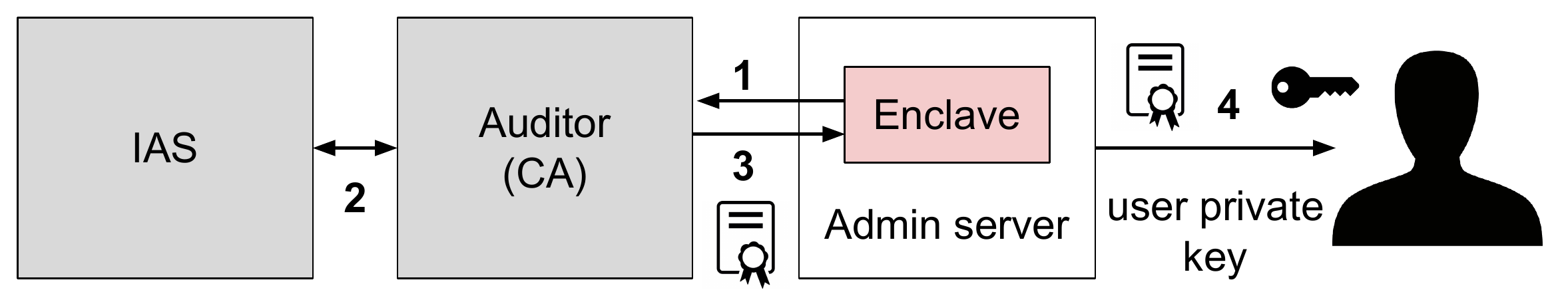}
    \caption{\label{fig:startup}Initial setup.}
\end{figure}

Figure~\ref{fig:startup} illustrates the initial setup of trust that must be
executed at least once before any key leaves the enclave.
Initially, the enclaved code generates a pair of asymmetric keys. 
While the private one never leaves the trusted domain, the public key is sent
along with the enclave measurement to the Auditor (1), who is both responsible 
for attesting the enclave and signing its certificate, thus also acting as a
\ac{CA}.
Next, the Auditor checks with \ac{IAS} (2) if the 
enclave is genuine.
Being the case, it compares the enclave measurement with the expected one, so 
that it can be sure that the code inside the shielded execution context is trustworthy.
Once that is achieved, the \ac{CA} issues the enclave's certificate (3), which also
contains its public key.
Finally, users are able to receive their private keys and the enclave's
certificate (4). 
The key will be encrypted by the enclave's private key generated in the 
beginning.
To be sure they are not communicating with rogue key issuers, users check
the signature in the certificate and then use the enclave's public key contained
within.
All communication channels described in this scheme must be encrypted by 
cryptographic protocols such as \ac{TLS}.

\subsection{From IBBE to IBBE-SGX}

Traditionally, the \ac{IBBE} scheme~\cite{sakai2007identity,Delerablee:2007:IBBE} consists of the following four operations.

\subsubsection{System Setup} The system setup  operation is run once by a \acf{TA} and generates a Master Secret Key $M_{SK}$ and a system-wide Public Key $PK$.

\subsubsection{Extract User Secret} The \ac{TA} then uses the Master Secret Key $M_{SK}$ to extract the secret key $U_{SK}$ for each user $U$.

\subsubsection{Encrypt  Broadcast Key} The broadcaster generates a randomized Broadcast Key $b_k$ for a given set of receivers $\mathcal{S}$, by making use of $PK$.
Together with $b_k$, the operation outputs a public broadcast ciphertext $c$.
The broadcast ciphertext can be publicly sent to members of $\mathcal{S}$ so they can derive $b_k$. 

\subsubsection{Decrypt Broadcast Key}  Any member of $\mathcal{S}$ can discover $b_k$ by performing the decrypt broadcast key operation given her secret key $U_{SK}$ and $\left(\mathcal{S}, c\right)$. 

\thinspace

In contrast to traditional \ac{IBBE} that requires the use of \iac{TA} to perform the \textit{System Setup} and \textit{Extract User Secret} operations, we rely on \ac{SGX} enclaves. 
Therefore, the master secret key \texttt{MSK} used by the two aforementioned operations can be made available in plaintext exclusively inside the enclave, and securely sealed if stored outside of the enclave for persistence reasons.

Similarly to \ac{IBBE}, the \textit{Encrypt Broadcast Key} and \textit{Decrypt Broadcast Key} operations  rely on the system public key $PK$, and are thus usable by any user of the system.

As opposed to the  traditional \ac{IBBE} usage scenario, our model  requires that all group membership changes---generating the group key and metadata---are performed by an \textit{administrator}. 
Administrators can use the master secret key \texttt{MSK} to encrypt, set up the system and extract user keys.
The decryption operation, however, remains identical to the traditional \ac{IBBE} approach, being executed by any arbitrary user.
In the remainder of this paper, we refer to our new \ac{IBBE} scheme as \textbf{IBBE-SGX}.

We now describe the computational simplification opportunities introduced by IBBE-SGX compared to \ac{IBBE}~\cite{Delerablee:2007:IBBE}.
First, by making use of \texttt{MSK} inside the enclave, the complexity of the encryption operation drops from $O(|S|^2)$ for \ac{IBBE} to $O(|S|)$ for IBBE-SGX, where $|S|$ is the number of users in the broadcast group set.
The reason behind the complexity drop is bypassing a polynomial expansion of quadratic cost, necessary in the traditional \ac{IBBE} assumptions.
The reader is directed to Section~\ref{sec:appendix:encrypt} for the concrete mathematical inference process.
We argue that this complexity cut is sufficient to tackle the impracticality of the \ac{IBBE} scheme emphasized earlier in Figure~\ref{fig:bb_test}.
Second, by relying on \texttt{MSK}, one can build efficient access control specific operations, such as adding or removing a user from a broadcast group.
IBBE-SGX can accommodate $O(1)$ complexities for both operations, as illustrated in Sections~\ref{sec:appendix:add} and~\ref{sec:appendix:remove}.

Unfortunately, IBBE-SGX maintains an $O(|S|^2)$ complexity for the user decrypt operation, during which, similarly to \ac{IBBE} encryption, the algorithm performs a polynomial expansion of quadratic cost. 
We address this drawback by introducing a partitioning mechanism as described later in Section~\ref{sec:partioning}.

Finally, we consider a re-keying operation, for optimally generating a new broadcast key and metadata when the identities of users in the group $\mathcal{S}$ do not change.
The operation can be performed in $O(1)$ complexity for both IBBE and IBBE-SGX, as detailed in Section~\ref{sec:appendix:rekey}.

\subsection{Partitioning Mechanism for IBBE-SGX }\label{sec:partioning}

Although IBBE-SGX produces a minimal metadata expansion and offers an optimal cost for group membership operations, it suffers from a prohibitive cost when a member needs to decrypt the broadcast key.
To address this issue, we introduce a partitioning mechanism. 

As the decryption time is bound to the number of users in the receiving set, we split the group into partitions (sub-groups) and therefore limit the user decryption time to the number of members in a single partition.
Moreover, each partition broadcast key will wrap the prime group key \texttt{gk}, so that members of different partitions can communicate by making use of \texttt{gk}.

\begin{figure}
	\centering
	\includegraphics[trim=0cm 0cm 0cm 3.5cm,width=0.4\textwidth]{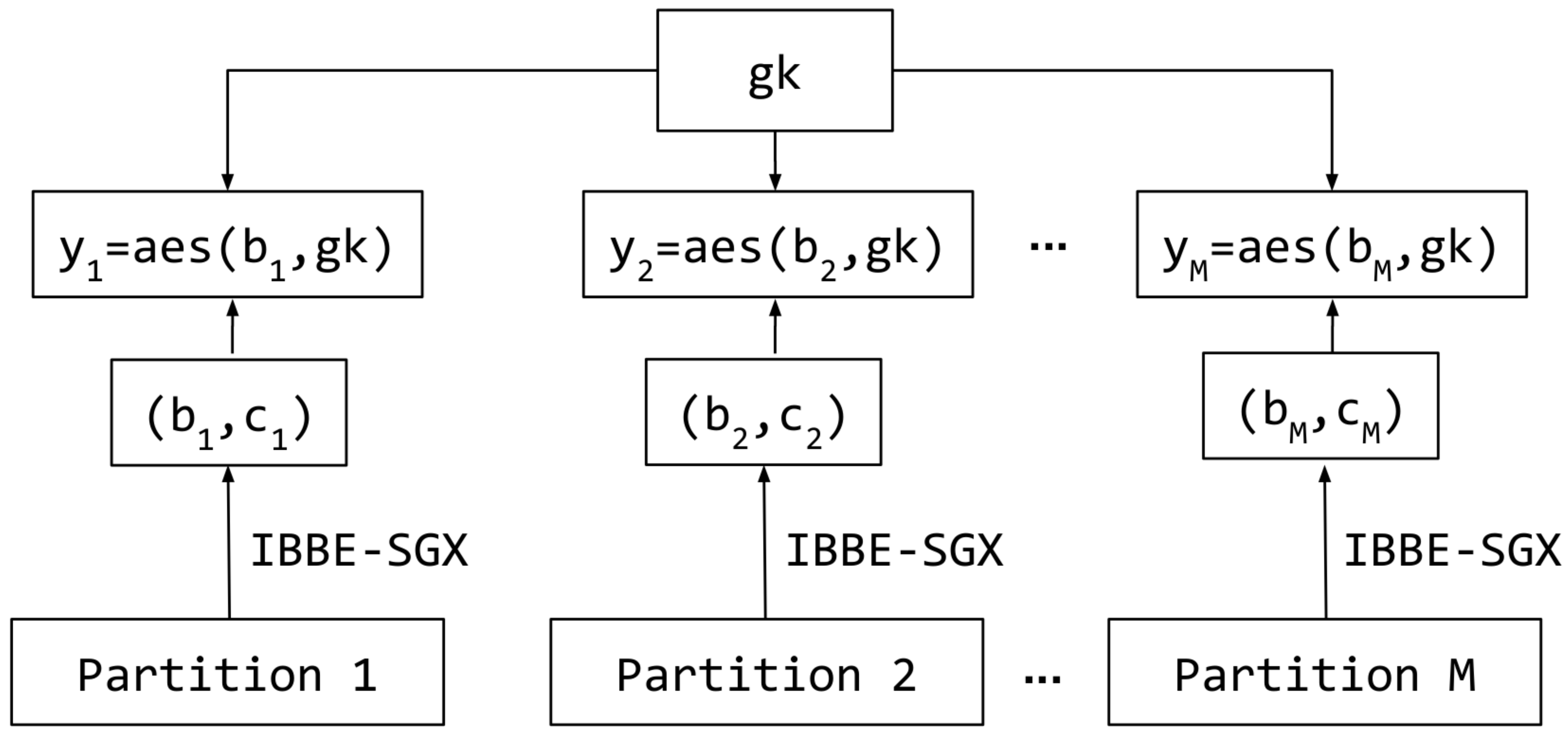}
	\caption{\label{fig:partitioning}Partitioning mechanism using IBBE-SGX and AES to protect the Group Key.}
\end{figure}

The partition mechanism is depicted in Fig~\ref{fig:partitioning}.  
The first step consists in splitting the group of users in fixed-size partitions.
The administrator can then use the encrypt functionality of IBBE-SGX  to generate a sub-group broadcast key $b_k$ and ciphertext $c_k$ for each partition $k$.
Next, for each partition, the group key \texttt{gk} is  encrypted using symmetric encryption such as \ac{AES}, by using the partition broadcast key $b_k$ as the symmetric encryption key.
Note that since the scheme is executing inside an SGX enclave, a curious administrator cannot observe \texttt{gk} nor the broadcast keys.

The group metadata of IBBE-SGX is therefore represented by the set of all pairs composed of the partition ciphertext and the encrypted group key (\textit{i.e.} ($c_i$, $y_i$) in Figure~\ref{fig:partitioning}). 
The inquisitive cloud storage can then publicly receive and store this set of group metadata.

Whenever a membership change happens, the administrator will update the list of group members and send the affected partition metadata to the cloud.
The clients, in turn, can detect a change in their group by listening to updates in their partition metadata.

The partitioning mechanism has an impact on the computational complexity of the IBBE-SGX scheme on the administrator side.
First, as the public key $PK$ of the IBBE system is linear in the maximal number of users in a group~\cite{Delerablee:2007:IBBE}, results that the public key for the  IBBE-SGX scheme is linear in the maximal number of users in a partition (denoted by $|p|$).
Therefore, both the computational complexity and storage footprint of the system setup phase can be reduced by a factor representing the maximal number of partitions, without losing any security guarantee.
Second, the complexities of IBBE-SGX operations change to accommodate the partitioning mechanism, as shown in Table~\ref{tab:complexity_drop}. 
Creating a group becomes the cost of creating as many IBBE-SGX partitions that the fixed partition size dictates.
Adding a user to a group remains constant, as the new user can be added either to an existing partition or to a brand new one.
Removing a user implies performing a constant time re-keying for each partition.
Finally, the decryption operation gains by being quadratic in the number of users of the partition rather than the whole group.

\begin{table}
	\caption{\label{tab:complexity_drop} IBBE-SGX and IBBE operations complexities per the number of partitions of a group ($|P|$), the fix size of a partition ($|p|$) and the cardinality of the group members set ($|S|$).}
	\centering
	\begin{tabular}{ llll }	
		\toprule
		Operation & IBBE-SGX & IBBE~\cite{Delerablee:2007:IBBE} \\
		\midrule
		System Setup  &$O(|p|)$ & $O(|S|)$ \\
		Extract User Key &$O(1)$ & $O(1)$ \\
		Create Group Key & $|P| \times O(|p|)$& $O(|S|^2)$ \\
		Add User to Group & $O(1)$ & \\
		Remove User from Group & $|P| \times O(1)$ & \\
		Decrypt Group Key &$O(|p|^2)$ & $O(|S|^2)$ \\
		\bottomrule
	\end{tabular}	
	\vspace{-0.65cm}
\end{table}

The partitioning mechanism also has an impact on the storage footprint for group metadata.
Compared to IBBE when considering a single partition, the footprint is augmented by the symmetrically encrypted partition broadcast key (i.e. $y_i$) and the \textit{nonce} required for this symmetric encryption.
When considering an entire group, the cost of storing the group metadata is represented by the cost of a single partition multiplied by the number of partitions in the group, in addition to a metadata structure that keeps the mapping between users and partitions.

Although the partition mechanism induces a slight overhead,  the number of partitions in a group is relatively small compared to the group size. 
Second, partition metadata are only manipulated by administrators, so they can locally cache it and thus bypass the cost of accessing the cloud for metadata structures.
Third, as our model accepts that the identities of group members can be discovered by the curious administrator or the cloud, there is no cryptographic operation needed to protect the mappings within the partition metadata structure.

Determining the optimal value for the partition size mainly depends on the dynamics of the group. 
Indeed, there is a trade-off between the number and frequency of operations performed by the administrator for group membership and those performed by regular users for decrypting the broadcast key.
A small partition size reduces the decryption time on the user side while a larger partition size reduces the number of operations performed by the administrator to run IBBE-SGX and to maintain the metadata.

\section{IBBE-SGX Group Access Control System}\label{sec:implementation}

\begin{figure}[t]
	\centering
	\includegraphics[width=0.4\textwidth]{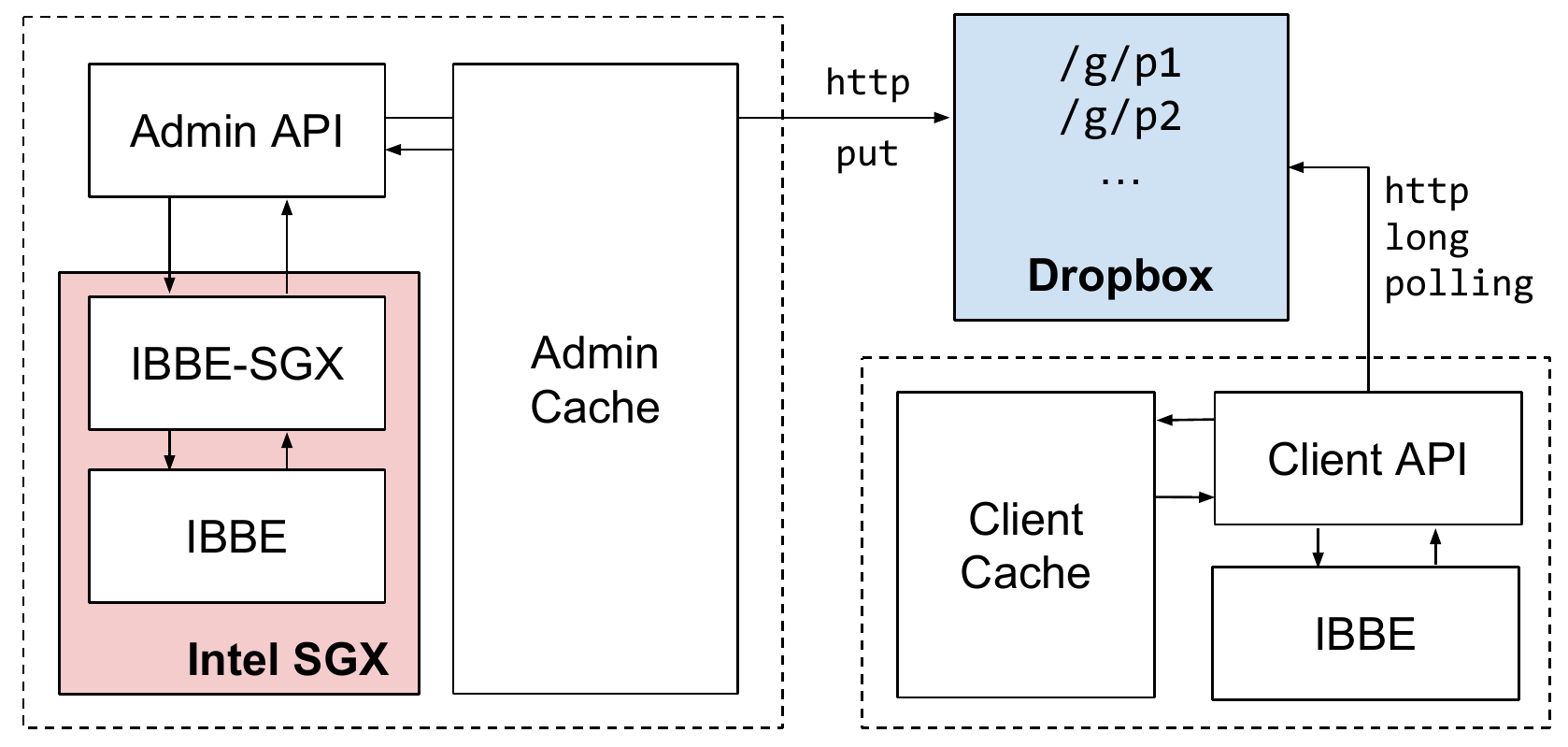}
	\caption{\label{fig:model}Big Picture Architecture}
\end{figure}

We describe in this section the design and implementation of an end-to-end group access control system based on IBBE-SGX. The overall architecture is illustrated in Figure~\ref{fig:model} and consists of a client and an administrator using Dropbox as a public cloud storage provider. 

\subsection{System Design}

The administrator's \ac{API}  makes calls to the underlying \ac{SGX} enclave that hold the functionalities of IBBE-SGX which is built on top of an \ac{IBBE} component. 
Since SGX is not required on the client side, the \textit{Client \ac{API}} directly calls the functionalities of the IBBE component.
Both administrators and clients make use of local in-memory caches in order to save round-trips to the cloud for accessing existing access policies.
Administrators make use of the \texttt{PUT} HTTP verb to send data to the cloud, while clients are listening by using \textit{HTTP long polling}.
In Dropbox, \textit{long polling} works at the directory level, so we index the group metadata as a bi-level hierarchy.
The parent folder represents the group, and each child stands for a partition.

The operation for creating a group is described in Algorithm~\ref{alg:create}. 
Once the fixed-size partitions are determined (line 1), the execution enters the SGX enclave (lines 2 to 6) during which the random group key is enveloped by the hash of each partition broadcast key.
The ciphertext values, as well as the sealed group key, leave the enclave to be later pushed to the cache and the cloud (line 7).

\setlength{\textfloatsep}{3pt}
\begin{algorithm}
	\caption{Create Group} %
	\textbf{Input:} Group $g$, Members $S = \{u_1, ... , u_n\}$, Partition size $m$
	\begin{algorithmic}[1] 		
		\State $\mathcal{P} \gets \{\{u_1,...,u_m\}, \{u_{m+1},...,u_{2m}\} ,... \}$%
		\begin{tcolorbox}
		\State $gk \gets RandomKey()$
			\For{$p \in \mathcal{P}$}
			\State $(b_p, c_p) \gets sgx\_ibbe\_create\_partition(\texttt{MSK}, p)$
			\State $y_p \gets sgx\_aes(sgx\_sha(b_p), gk)$
			\EndFor
			\State $sealed\_gk \gets sgx\_seal(gk)$
        \end{tcolorbox}
		\State Store: (1) $sealed\_gk$; (2) $\forall p \in P : \langle \forall u \in p, y_p, c_p \rangle$
	\end{algorithmic}
	\label{alg:create}
	
\end{algorithm}

The operation of adding a user to a group (Algorithm~\ref{alg:add}) starts by finding the set of all partitions with remaining capacity (line 1).
If no such a partition is found, a new partition is created for the user (line 3) and the group key is enveloped by the broadcast key of the new partition (lines 4 to 6), before persisting its ciphertexts (line 7).
Otherwise, a partition that is not empty is randomly picked, and the user is added to it (lines 9, 10).
Since the partition broadcast key remains unchanged, only the ciphertext needs to be adapted to include the new user (line 10).
The partition members and ciphertext are then updated on the cloud (line 12).
Note that there is no need to push the encrypted group key $y_{add}$ as it was not changed.
\vspace{-0.2cm}

\begin{algorithm}
	\caption{Add User to Group}
	\textbf{Input:} Group: $g$, Partitions of $g$: $\mathcal{P}$, User to add: $u_{add}$, Sealed group key: $sealed\_gk$. 
	\begin{algorithmic}[1]
		\State $\mathcal{P'} \gets \forall p \in \mathcal{P},\textnormal{ $such$ $that$ } \vert p \vert < m.$
		
		\If{$\mathcal{P'} = \emptyset$}
		\State $p_{add} \gets \{u_{add}\} $
		\begin{tcolorbox}
		\State $(b_{add}, c_{add}) \gets sgx\_ibbe\_create\_partition(\texttt{MSK}, p_{add})$
			\State $gk\gets sgx\_unseal(sealed\_gk)$
			\State $y_{add} \gets sgx\_aes(sgx\_sha(b_{add}), gk)$
        \end{tcolorbox}
		\State Store:
		$\langle \{u_{add}\}, y_{add}, c_{add} \rangle$
		\Else
		\State $p_{add} \gets RandomItem(\mathcal{P'}) $
		\State $p_{add} \gets p_u \cup \{u_{add}\} $
		\begin{tcolorbox}[notitle]
       		\State $c_{add}\gets sgx\_add\_user\_to\_partition(\texttt{MSK}, p_{add}, u_{add})$
        \end{tcolorbox}
		\State Update: $\langle \forall u \in p_{add}, * , c_{add} \rangle$
		\EndIf
		\State $\mathcal{P} \gets p_{add} \cup \mathcal{P}$
	\end{algorithmic}
	\label{alg:add}
\end{algorithm}
\vspace{-0.3cm}

Removing a user from a group (Algorithm~\ref{alg:remove}) proceeds by removing the user from her hosting partition (lines 1 and 2). 
Next, a new group key is randomly generated (line 3).
The former user hosting partition broadcast key and ciphertext are changed to reflect the user removal (line 4) and then used for enveloping the new group key (line 5).
For all the remaining partitions, a constant time re-keying regenerates the partition broadcast key and ciphertext that envelopes the new group key (lines 6 to 9).
After sealing the new group key (line 10), the changes of metadata for the group partitions are pushed to the cloud (line 11).
Note that the partition members only need to be updated for the removed user hosting partition.

\begin{algorithm}
	\caption{Remove User $u_{rem}$ from Group $g$}
	\textbf{Input:} Group: $g$, Partitions of $g$: $\mathcal{P}$, User to remove: $u_{rem}$.
	
	\begin{algorithmic}[1] 
		\State $p_{rem} \gets p \in \mathcal{P}, \ such \ that \ u_{rem} \in p$.
		\State $p_{rem} \gets p_{rem} \setminus \{u_{rem}\} $
		\begin{tcolorbox}
		\State $gk \gets RandomKey()$
			\State $(b_{rem}, c_{rem}) \gets sgx\_remove\_user(\texttt{MSK}, p_{rem}, u_{rem})$
			\State $y_{rem} \gets sgx\_aes(sgx\_sha(b_{rem}), gk)$
			\For{$p \in \mathcal{P} \setminus p_{rem}$}
			\State $(b_{p}, c_{p}) \gets sgx\_rekey\_partition(p)$
			\State $y_{p} \gets sgx\_aes(sgx\_sha(b_{p}), gk)$
			\EndFor		
			\State $sealed\_gk \gets sgx\_seal(gk)$
		\end{tcolorbox}
		\State Update: (1) $\langle \forall u_i \in p_{rem}, y_{rem}, c_{rem} \rangle$\\~~~~~~~~~ (2) $\forall p \in \mathcal{P} \setminus p_{rem}$ : $\langle * , y_{p}, c_{p} \rangle$
	\end{algorithmic}
	\label{alg:remove}
\end{algorithm}

As many removal operations can result in partially unoccupied partitions, we propose the use of a re-partitioning scheme whenever the partition occupancies are too low.
We implement a heuristic to detect a low occupancy factor such that if less than half of the partitions are only two thirds full, then re-partitioning is triggered.
Re-partitioning consists in simply re-creating the group following Algorithm~\ref{alg:create}.

Finally, the client decrypt operation works by first using IBBE to decrypt the broadcast key and then use the hash of this key for an \ac{AES} decryption to obtain the group key. 
Due to space constraints, we omit the formal algorithm specification.

\subsection{Implementation}

In order to implement the system, we used the PBC~\cite{lynn2006pbc} \emph{pairing-based cryptography} library which, in turn, depends on GMP~\cite{granlund1991gmp} to perform arbitrary precision arithmetics.
They both have to be used inside SGX enclaves (Section~\ref{sec:solution}).
There are several challenges when porting legacy code to run inside enclaves.
Besides having severe memory limitations (Section~\ref{subsec:backgroundsgx}), it also considers privileged code
running in any protection ring but user-mode (ring 3) as not trusted.
Therefore, enclaves cannot call operating system routines.

Although memory limitations can have performance implications at runtime, they
have little influence on enclave code porting.
Calls to the operating system, on the other hand, can render this task very
complex or even unfeasible.
Luckily, since both PBC and GMP mostly perform computations rather than input
and output operations, the challenges on adapting them were chiefly restrained
to tracking and adapting calls to \emph{glibc}.
The adaptations needed were done either by relaying operations to the operating
system through \emph{outside calls} (ocalls), or 
performing them with enclaved equivalents.
The outside calls, however, do not perform any sensitive action that could compromise security.
Aside from source code modifications, we dedicated efforts to adapt the
compilation toolchain. This happens because one has to use curated versions
of standard libraries (like the ones provided by Intel SGX SDK), 
besides having to prevent the use of compiler's built-in functions and
setting some other code generation flags.
The total number of \acp{LoC} or compilation toolchain files that
were modified were 32 lines for PBC and 299 for GMP.

Apart from changes imposed by SGX, we also needed to use common cryptographic libraries. 
Although some functions are provided in v.1.9 of the Intel~SGX~SDK~\cite{sgx-sdk}, its \ac{AES} implementation is limited to 128 bits.
Since we aim at the maximal security level, we used the \ac{AES} 256 bits implementation provided in Intel's port of OpenSSL~\cite{intel-sgx-ssl}.
The end-to-end system encapsulating both IBBE-SGX and \ac{HE} schemes consists in 3,152 lines of C/C++ code and 170 lines of Python.
\section{Evaluation}
\label{sec:evaluation}

In this section, we benchmark the performance of the IBBE-SGX scheme from three different perspectives: by measuring the operations performance in isolation, then by comparing them to \acf{HE}, and finally by capturing the performance when replaying realistic and generated access control traces.
We chose to compare IBBE-SGX to HE only as the latter already shows better computational complexity than IBBE (see Figure~\ref{fig:bb_lat}).

The experiments were performed on a quad-core Intel i7-6600U machine, having a processor at 3.4 GHz with 16 GB of RAM, using Ubuntu 16.04 LTS.

\subsection{Microbenchmarks}
Within the microbenchmarks we isolate the performance of each IBBE-SGX operation, and perform a comparison with the \ac{HE} scheme.

\begin{figure}[!t]
	\centering
	\subfloat[System Set-Up Latency]{\includegraphics[width=0.5\linewidth]{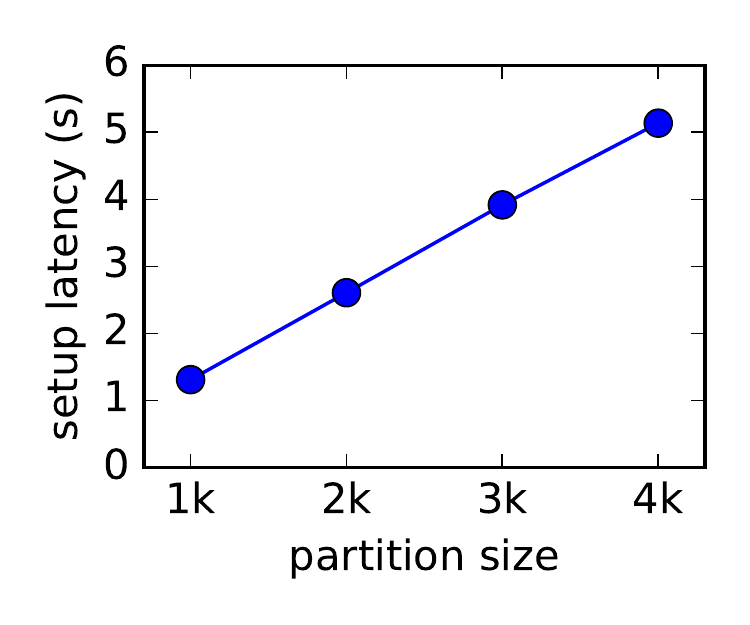}%
		\label{micro_setup}}
	\hfil
	\subfloat[Key Extract Throughput]{\includegraphics[width=0.5\linewidth]{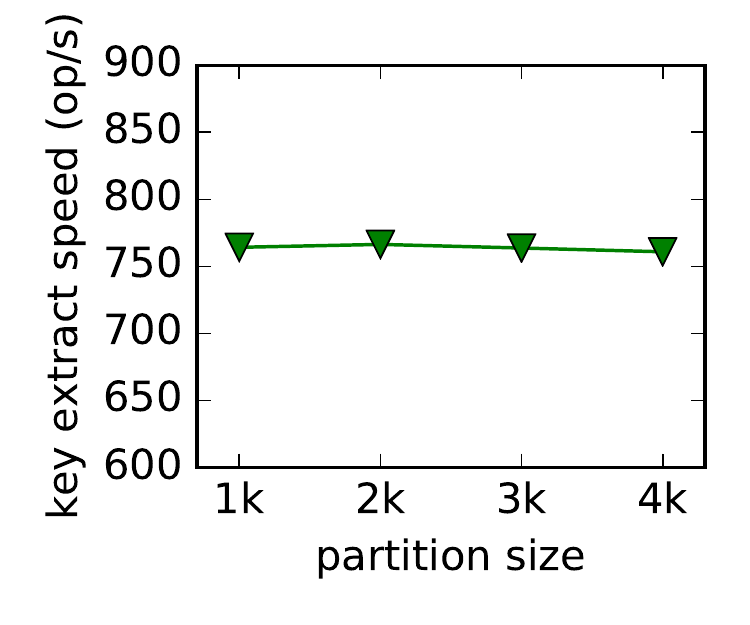}%
		\label{micro_key}}
	\caption{Performance of the system bootstrap phase.}
	\label{fig:micro_bootstrap}
\end{figure}

First, we evaluate the performance of the bootstrap phase. It consists on setting up the system and generating secret user keys, referenced in Figure~\ref{fig:micro_bootstrap}. 
One can notice that the setup phase latency increases linearly per partition size, with a growth of 1.2s per 1,000 users.
In contrast, extracting secret user keys gives an average throughput of 764 operations per second, independent of the partition size.

\begin{figure}[!t]
	\centering
	
	\subfloat{\includegraphics[width=0.8\columnwidth]{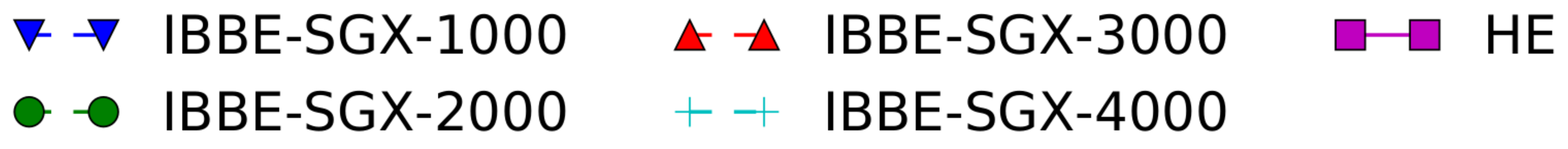}}
	\\[-2ex]
	\addtocounter{subfigure}{-1}
	\subfloat[Comparing IBBE-SGX and HE]{\includegraphics[width=0.5\columnwidth]{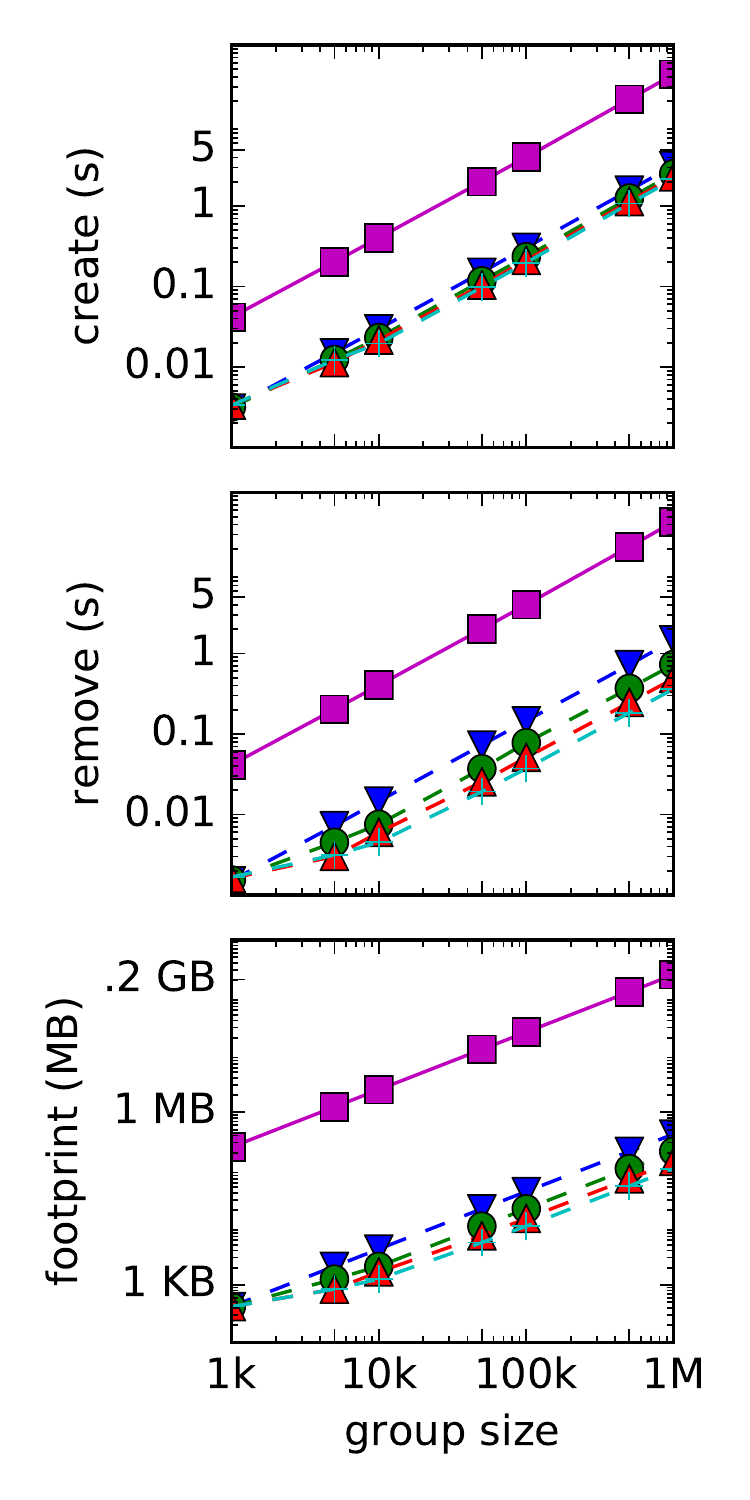}%
		\label{fig:micro_loglog}}
	\hfil
	\subfloat[Partitioning Evaluation]{\includegraphics[width=0.5\linewidth]{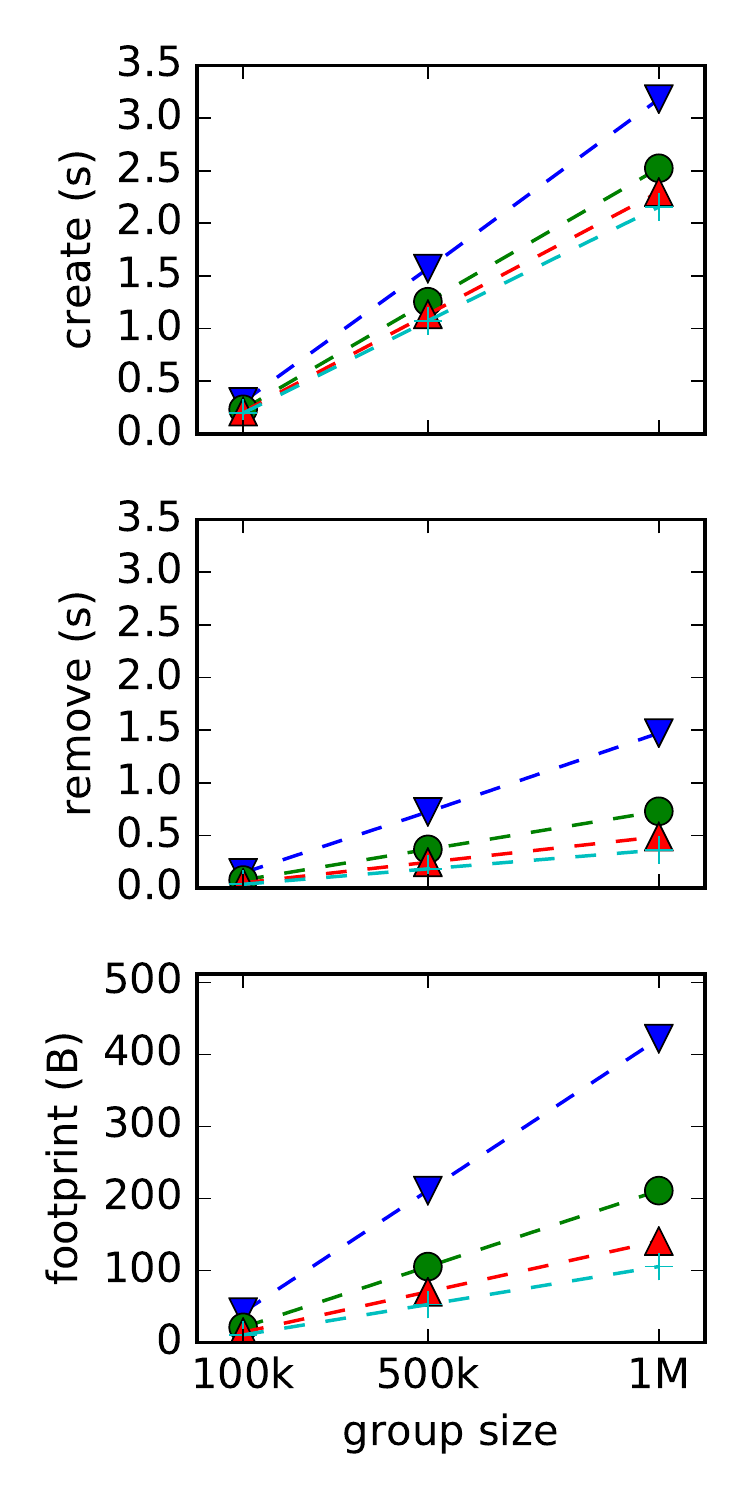}%
		\label{fig:micro_lin}}

	\caption{Evaluation of create and remove operations and storage footprint, by varying the partition size for IBBE-SGX (1000, 2000, 3000 and 4000).}
	\label{fig:}
\end{figure}

Next, we evaluate the behavior of IBBE-SGX operations compared to \ac{HE}.
Figure~\ref{fig:micro_loglog} displays the computational cost for operations of creating a group, removing a user from a group, and the storage footprint of the group metadata.
One can notice that all three operations are better than their \ac{HE} counterparts by approximately a constant factor.
The computational cost of create and remove operations of IBBE-SGX is on average 1.2 orders of magnitude faster than \ac{HE}.
Compared to the original IBBE scheme, IBBE-SGX is better by 2.4 orders of magnitude for groups of 1,000 users and 3.9 orders of magnitude for one million users (see Figures~\ref{fig:bb_lat} and~\ref{fig:micro_loglog}).
Storage-wise, IBBE-SGX is up to 6 orders of magnitude better than \ac{HE}.
Moreover, Figure~\ref{fig:micro_lin} zooms into the performances of IBBE-SGX create and remove operations, and the storage footprint respectively, when considering different sizes of partitions.
One can notice that the remove operation takes half the time than the create operation.
Considering the storage footprint, the degradation brought by using smaller partition sizes is fairly small (e.g., 432 vs. 128 bytes for groups of 1 million members).

\begin{figure}[!t]
	\centering
	
	\subfloat{\includegraphics[width=0.34\columnwidth]{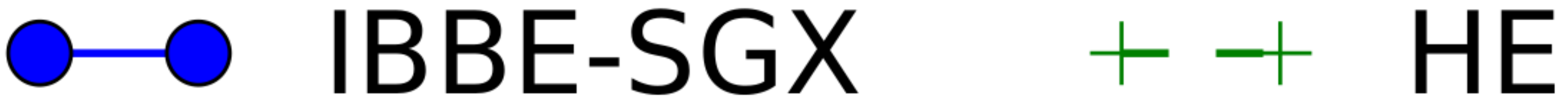}}
	\\[-2ex]
	\addtocounter{subfigure}{-1}
	\subfloat[Add user to group performance.]{\includegraphics[width=0.5\linewidth]{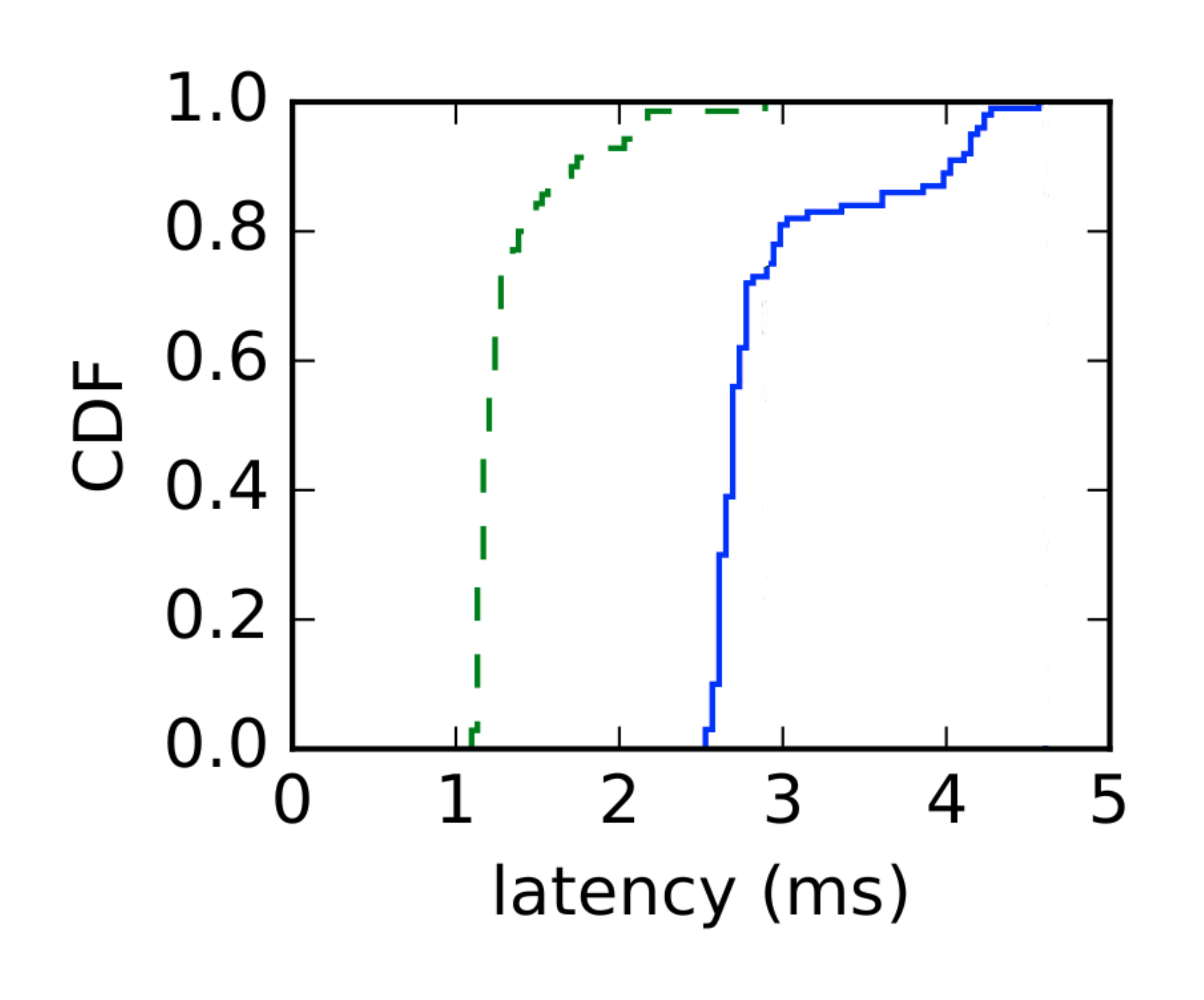}%
		\label{fig:add}}
	\hfil
	\subfloat[Decrypt performance.]{\includegraphics[width=0.5\linewidth]{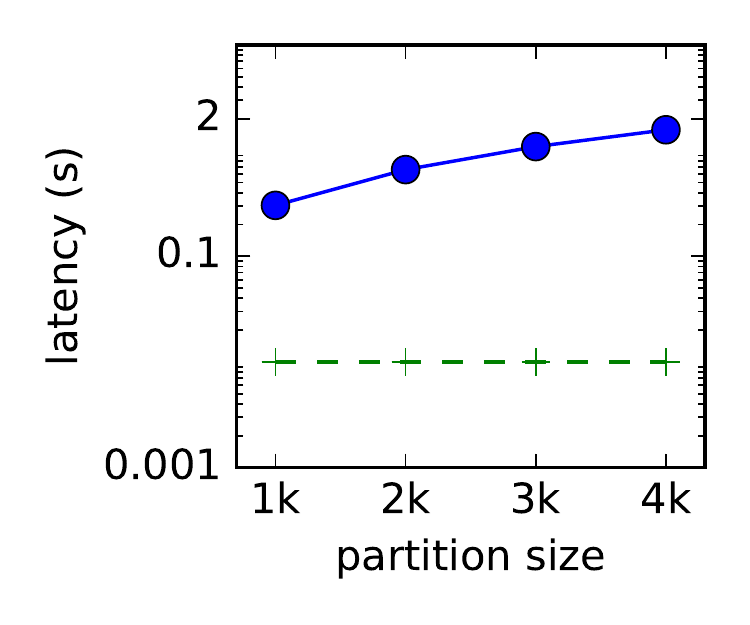}%
		\label{fig:decrypt}}
	\caption{Performance of the adding a user to a group and decrypt operations.}
	\label{fig:perf_final}
\end{figure}

The \ac{CDF} of latencies for adding a user to a group is shown in Figure~\ref{fig:add}.
The operation has a constant time complexity for both IBBE-SGX and HE.
As the add operation of IBBE-SGX can take two paths, either adding a user to an existing partition or creating a new one if all the others are full, the plot points the difference between the two at the \ac{CDF} value of 0.8.
Moreover, the \ac{HE} add operation is generally twice as fast as IBBE-SGX.
 
The client decryption performance is shown in Figure~\ref{fig:decrypt}.
The decryption operation, like the add operation, is faster within the \ac{HE} approach than IBBE-SGX.
The difference of 2 orders of magnitude is caused by the quadratic cost of the IBBE-SGX decryption operation.
We argue that a slower decryption time for IBBE-SGX can be acceptable in practice.
First, the decrypt performance is overshadowed by the slow cloud response time necessary for clients to update the group metadata that always precedes a decryption operation.
Second, the cost of decryption remains bounded to a partition size, independent on the number of users in the group.
\subsection{Macrobenchmarks}

\subsubsection{Real Data Set}
\begin{figure}[!t]
	\centering
		\vspace{-0.5cm}
		\subfloat{\includegraphics[width=0.48\columnwidth]{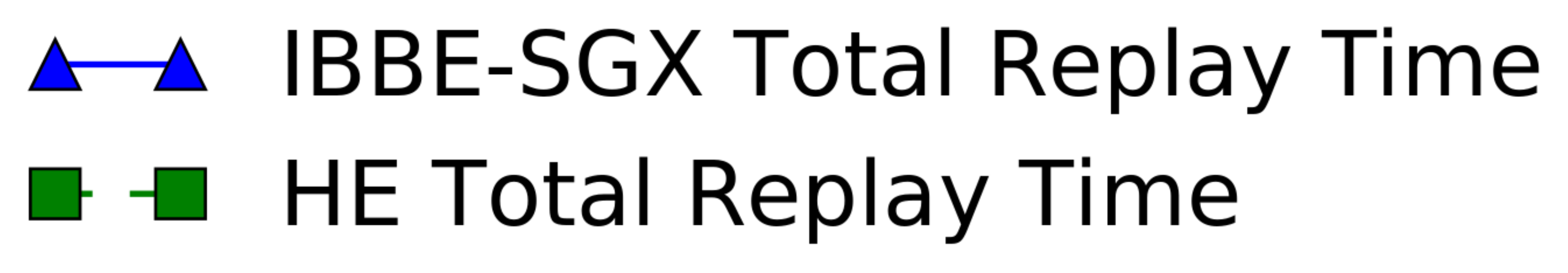}}
		\hfill
		\subfloat{\includegraphics[width=0.48\columnwidth]{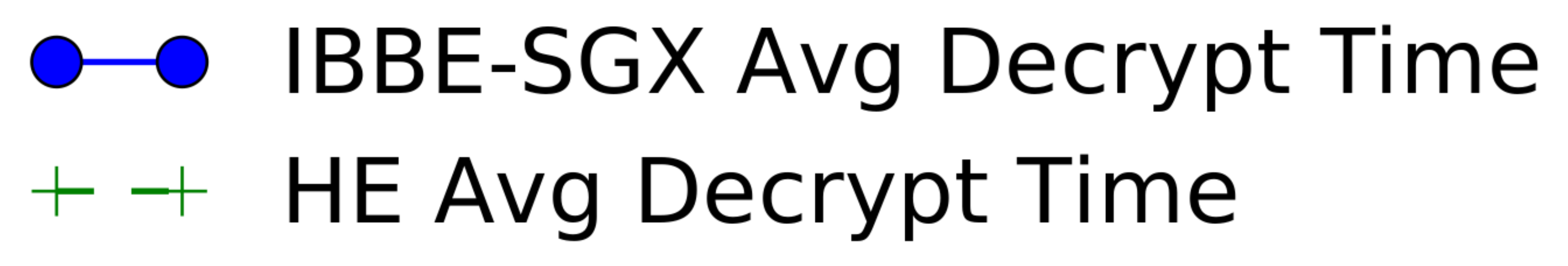}}
		
	{\includegraphics[width=.9\linewidth]{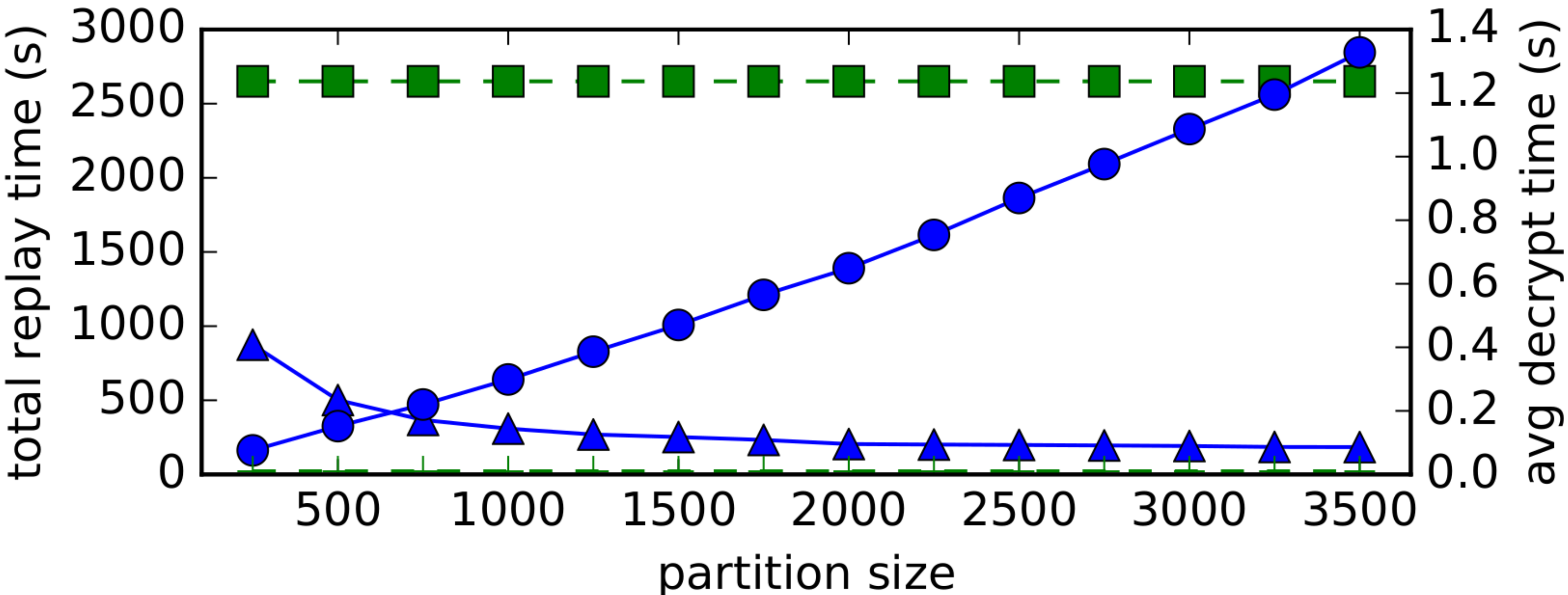}%
}
	\caption{Measuring total administrator replay time and average  user decryption time per different partition sizes using the Linux Kernel ACL data set.}
	\label{fig:macro_linux}
\end{figure}

To capture the performance of the IBBE-SGX scheme within a realistic scenario, we decide to replay an access control trace based on the membership changes in the version control repository of the Linux Kernel~\cite{_linux_dataset}.

We derive the membership trace by considering the first commit of a user as the add to group operation. 
The remove from group operation is represented by the user's last commit. 
The generated trace contains 43,468 membership operations that spawn across a period of 10 years, during which the group size never exceeds 2803 users.
We replay the generated trace sequentially for both \ac{HE} and IBBE-SGX by varying the partition size. We also capture the total time spent by the administrator to replay the trace and the average user decryption time.
 
Figure~\ref{fig:macro_linux} displays the results.
Considering the administrator replay time, IBBE-SGX performs better when the partition size converges to the number of users in the group. 
Using a small partition size, e.g. 250, is almost twice as inefficient when compared to a partition of 1000 users.
Compared to HE replay time, IBBE-SGX is generally 1 order of magnitude faster. 
On the other hand, decryption time for IBBE-SGX grows quadratically per partition size while in HE it remains constant.
This evidentiates IBBE-SGX's trade-off caused by different partition sizes on the performances of membership changes and user decryption time.
A prior estimation of the maximal group size (2803 in our case) suggests the choice of a small partition for practical use (such as 750), so that it can manifest satisfactory outcomes both in terms of admininistrator performance and user decryption time.

\subsubsection{Synthetic Data Set}
In order to observe the impact of different workloads of group membership access control, we generate a set of synthetic traces that capture incremental percentages of revocation rates.
Concretely, we generate 11 different traces consisting of 10,000 membership operations.
The composition of the traces is randomly generated by considering different revocation rates.
We replay the 11 traces through our system and measure the end-to-end time required by the administrator to perform all membership changes.
We then repeat the process by considering different partition sizes.

\begin{figure}[!t]
	\centering
	
	\subfloat{\includegraphics[width=0.9\columnwidth]{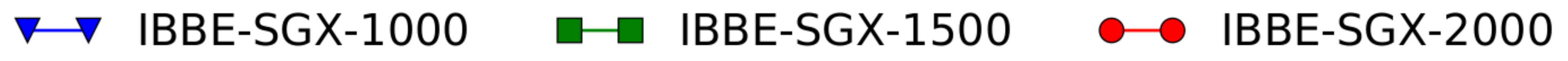}}
	
	{\includegraphics[width=0.9\linewidth]{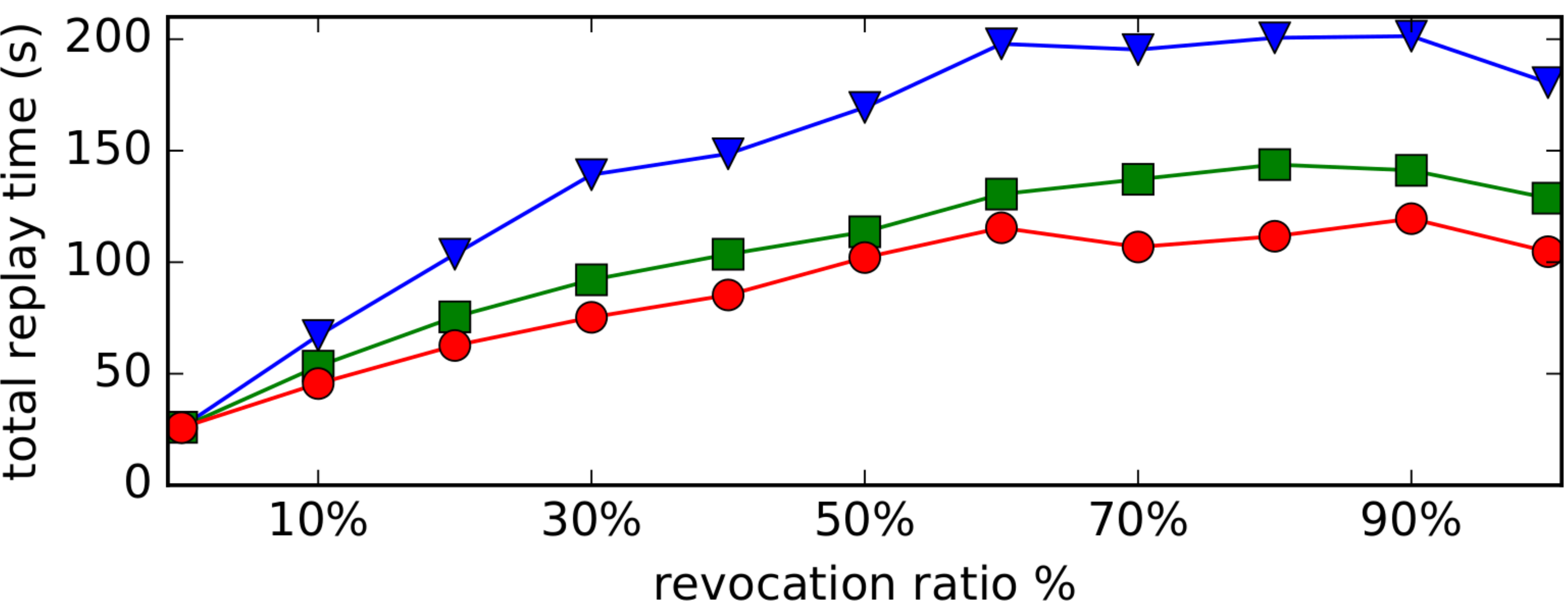}%
		\label{macro_rev}}
	\caption{Measuring total replay time of IBBE-SGX scheme per different partition sizes (1000, 1500, 2000) for synthetically generated datasets considering increasing revocation rates.}
	\label{fig:macro_syn}
\end{figure}

The results are shown in Figure~\ref{fig:macro_syn}.
We observe a linear increase in the total time when incrementally increasing the revocation ratio up to 50\% in workloads dominated by add operations. 
After this point, the total time stabilizes and finally decreases when the revocation ratio is more than 90\%. 
This behavior is caused by the merging of sparse partitions, which happens more frequently with the increasing rate of revocations.
Having fewer partitions, IBBE-SGX's operations become faster, therefore decreasing the total time.

\section{Related Work}
\label{sec:related_work}

We structure the presentation of the related work on three axes detailing first research work on cryptographic schemes used for access control.
Then we go into research work of systems that cryptographically protect from untrusted storages.
Last, we detail related work regarding Intel SGX.

\subsection{Cryptographic Schemes for Access Control}

\ac{HE} making use of a \ac{PKI} and \ac{IBE} has been utilized within a role based access control and proved unsuitable in the cloud storage context~\cite{Garrison:2016:DACCloud}.

\ac{ABE}~\cite{sahai2005fuzzy} is a cryptographic construction that allows a fine-grained access control by matching attributes labeled to both users and content.
Depending on the labeled location, one can distinguish between key-policy \ac{ABE}~\cite{goyal2006attribute} and ciphertext-policy \ac{ABE}~\cite{bethencourt2007ciphertext}.
However, when employed for simple access control policies, such as our group sharing context, \ac{ABE} has substantially greater costs than identity-based encryption~\cite{Garrison:2016:DACCloud}.

\ac{HIBE}~\cite{boneh2005hierarchical} and \ac{FE}~\cite{boneh2011functional} are two cryptographic schemes offering functionalities for access control that, similarly to \ac{IBE} and \ac{ABE}, rely on pairing-based cryptography. 
\ac{HIBE} is specifically designed to target hierarchical organizations where a notion of descendants exists.
\ac{FE} is a powerful construction that can arbitrarily encapsulate programs as access control, but is unsuitable for practical use~\cite{fischiron}.

Proxy re-encryption~\cite{ateniese2006improved} is a cryptographic system in which the owner of some encrypted data can delegate the re-encryption of her data to a proxy, with the intent of sharing it with another user.
For the re-encryption to take place, the data owner needs to generate and transmit to the proxy a re-encryption key.
The scheme proves to be beneficial for the cloud environment, as the re-encryption and the storage of the data can happen on the same premises.
A number of approaches have shown how proxy re-encryption can be combined with identity-based encryption ~\cite{green2007identity}, or with attribute-based encryption ~\cite{green2011outsourcing, sahai2012dynamic}.
Differently than proxy re-encryption, our construction does not require users to send transformational keys to the administrators.
Therefore, even if the administrators would be hosted on the cloud storage premises, they do not act as proxies.

The related research area of \textbf{\textit{multicast communication security}}~\cite{stinson2005cryptography, canetti1999multicast} defines efficient schemes focusing exclusively on revocation aspects.
Logical Key Hierarchy~\cite{wallner1999key} is a re-keying scheme in which communications for revocation operations are minimized to logarithmic sizes.
Other schemes~\cite{fiat1993broadcast, naor2000efficient} exploit a secret sharing mechanism, considering that no coalition of revoked users larger than a threshold number is trying to decrypt the transmission.

\subsection{Cryptographically Protected Untrusted Storages}

The shared cloud-backed file system (SCFS) designed by \textit{Bessani et al.}~\cite{bessani2014scfs} offers confidentiality guarantees to users by encrypting data stored by the clouds on the client-side.
Even though the encryption keys are distributed among multiple cloud storages through secret sharing schemes, the access control is not cryptographically protected, but stored and enforced from a trusted coordination service.
We argue that this approach is not secure enough  because it does not protect from curious administrators.
The global access control structure can be compromised if an attacker gain access to this service.

CloudProof~\cite{popa2011enabling} is a secure cloud storage system offering guarantees such as confidentiality, integrity, freshness and write-serializability. 
To enforce access control, CloudProof makes use of broadcast encryption to protect the keys that are used for encrypting and signing the actual data. 
Unlike our construction, CloudProof does not offer the \emph{zero knowledge} guarantee for membership operations.
Moreover, CloudProof does not discuss how the authentic identity of the users in the broadcast set is established (\textit{e.g.} during a group creation operation).
Hypothetically, a \ac{PKI} could be employed for this task, thus requiring a trusted entity in the system.
However accessing regularly the \ac{PKI}  would add a significant overhead.
In order to mitigate these issues, our solution relies on the identity-based version of broadcast encryption.

REED~\cite{li2016rekeying} considers the problem of rekeying in the context of honest-but-curious deduplicated storages.
To provide access control, REED relies on \ac{ABE}.
However, as noted by the authors in their evaluation, the performance overhead of the rekeying operation drastically increase to several seconds when varying the total number of users up to 500.
Considering group sizes of thousands of users (as we do for groups up to one million), \ac{ABE} becomes impracticable for access control at large scale.

Sieve~\cite{wang2016sieve} platform allows users to store their data encrypted in the cloud and then discretionary delegate access to the data to consuming web services.
Sieve makes use of attribute based encryption for access control policies and key homomorphism for providing a \emph{zero knowledge} guarantee against the storage provider.
This access control construction has many similarities with ours, however we differentiate exploiting the \emph{zero knowledge} guarantee for lowering the computational complexity of the access control scheme, IBBE in our case.

\subsection{SGX}

SGX has been extensively used in shielding applications and infrastructure 
platform services like ours that handle sensitive data.
Iron \cite{fischiron} is the closest to our proposal in the sense that
it takes advantage of SGX to build a practical encryption scheme for an 
unpractical strategy thus far.
Like us, they use an enclave that holds a master secret as root for later key 
derivations.
They target, however, functional encryption. 
The enclave generates a key that is associated to a function, so that the
computation can be performed without revealing the data on top of which it is 
applied.
The results of applying such function, though, are presented in clear.
The authors show that this approach outperforms by orders of magnitude other 
cryptographic schemes that also offer functional encryption.

Other systems relate to ours with regards to the reduction of overhead for 
an otherwise costlier design.
Hybster \cite{Behl:2017:Hybrids}, for instance, proposes a hybrid state-machine
replication protocol. 
Hybrid because it does tolerate arbitrary faults but yet
it assumes that some nodes may crash. 
It relies on SGX features such as
isolation, replay protection and trusted counters to achieve a parallelization
scheme that makes it a viable solution for demanding applications, reaching
higher numbers of operations per second in comparison to traditional approaches.

At the level of infrastructure services, SCBR~\cite{Pires:2016:SCBR} proposes
a content-based routing solution where the filtering step is put inside
enclaves, thus allowing the matching of publications against stored
subscriptions in a safe manner. 
It is shown to be one order of magnitude faster than an approach with 
comparable security guarantees.
The gain comes from the plaintext operations done inside the enclave against
the counterpart that needs to perform computations over encrypted data.

\section{Conclusion}
\label{sec:conclusion}
\acresetall

We have introduced IBBE-SGX, a new cryptographic access control extension that is built upon \ac{IBBE} and exploits Intel \ac{SGX} to derive cuts in the computational complexity of \ac{IBBE}.
We propose a group partitioning mechanism such that the computational cost of membership update is bound to a fixed constant partition size rather than the size of the whole group.
We have implemented and evaluated our new access control extension in a single administrator with multiple users set-up.
We have conducted both real and synthetic benchmarks, demonstrating that IBBE-SGX is efficient both in terms of computation and storage even when processing large and dynamic workloads of membership operations.
Our innovative construction performs membership changes 1.2 orders of magnitude faster than the traditional approach of \ac{HE}, producing group metadata that are 6 orders of magnitude smaller than \ac{HE}, while at the same time offering \emph{zero knowledge} guarantees.

There are a number of interesting avenues of future work.
The first is to dynamically adapt the partition sizes based on the undergoing workload.
This would optimize the speed of administrator- and user-performed operations.
A second challenge would be to adapt our construction to a distributed set of administrators that would perform membership changes concurrently on the same group or partition, by using lock-free techniques.
Third, in a setup with multiple administrators, one can envision certifying blocks of membership operations logs through blockchain-like technologies.

\section*{Acknowledgment}
The research leading to these results has received funding from the French Directorate General of Armaments (DGA) under contract RAPID-172906010.
The work was also supported by
European Commission, Information and Communication Technologies, H2020-ICT-2015 under grant agreement number 690111 (SecureCloud project)
and partially supported by the CHIST-ERA "DIONASYS" project.

{
	\printbibliography
}

\appendices
\newpage
\section{}\label{sec:math}

\setlength{\abovedisplayskip}{0pt}
\setlength{\belowdisplayskip}{5pt}
\setlength{\abovedisplayshortskip}{0pt}
\setlength{\belowdisplayshortskip}{5pt}

This appendix details the mathematical implications of adapting IBBE to IBBE-SGX.
In typical IBBE schemes, \acp{TA} only execute the operations of system setup and extracting user secret keys.
In IBBE-SGX, however, the administrator agent executes all membership operations by executing them inside an \ac{SGX} enclave.

The IBBE scheme~\cite{Delerablee:2007:IBBE} conceptually relies on the idea of bilinear maps. Notated as: 
$ e\left(\cdot,\cdot\right) : \mathbb{G}_1 \times \mathbb{G}_2 \rightarrow \mathbb{G}_T $,
a bilinear map is defined by using three cyclic groups of prime order $p$, imposing bilinearity and non-degeneracy. 
\textit{El Mrabet et al.}~\cite{el2017guide} provide a thorough overview of bilinear maps usage within the cryptographic setting.
Moreover, the IBBE scheme implies the public knowledge of a cryptographic hash function $\mathcal{H}$, that maps user identity strings to values in $\mathbb{Z}_p^*$.

\subsection{System Setup}\label{sec:appendix:setup}
The initial operation is identical for IBBE and IBBE-SGX.
The algorithm receives $\left(\lambda, m\right)$ as input, where $\lambda$ represents the security strength level of the cryptosystem, and $m$ encapsulates the largest envisioned group size.
The output consists of the Master Secret Key $M_{SK}$ and the system Public Key $PK$. To build $M_{SK}$, the algorithm randomly picks $g \in \mathbb{G}_1 \text{and} \gamma \in \mathbb{Z}_p^*$ : $M_{SK} = \left(g, \gamma\right)$.
To construct the $PK$, the algorithm computes $w = g^\gamma$ and $v = e\left(g, h\right)$, where $h \in \mathbb{G}_2$ was randomly picked:
$ PK = \left(w, v, h, h^\gamma, h^{\gamma^2}, ... , h^{\gamma^m}\right)$.
The computational complexity of the system setup algorithm is linear to  $m$.

\subsection{User Key Extraction}\label{sec:appendix:extract}
The key extraction operation is identical for IBBE and IBBE-SGX.
For a given user identity $u$, the operation makes use of $M_{SK}$ and computes : 
$U_{SK}~=~g^{\left(\gamma + \mathcal{H}\left(u\right)\right)^{-1}}$.

\subsection{Encrypt Broadcast Key}\label{sec:appendix:encrypt}

The algorithm for constructing a broadcast key differs by considering the specific usage assumption. 
If for IBBE the algorithm has to rely on $PK$, for IBBE-SGX one can make use of $M_{SK}$.
In both cases, the group broadcast key is randomly generated by choosing a random value $k \in \mathbb{Z}_p^*$ and computing:

\begin{equation}
 b_k = v^k 
 \label{K_from_k}
\end{equation}
 
A group broadcast ciphertext $\left(C_1, C_2\right)$ is then constructed by:
\begin{align}
C_1 &= w^{-k} \label{c_1} \\
C_2 &= h^{k \cdot \prod\limits_{u \in S}\left(\gamma + \mathcal{H}\left(u\right)\right)} \label{c_2}
\end{align}

For IBBE, $\gamma$ cannot be used directly for computing $C_2$.
Instead, the computation is carried out with a polynomial expansion of the exponent that uses the public key elements:

\begin{equation}
C_2 =
\left( \left(h^{\gamma^n}\right)\cdot\left(h^{\gamma^{n-1}}\right)^{\mathcal{E}_1}\cdot\left(h^{\gamma^{n-2}}\right)^{\mathcal{E}_2}\cdot ... \cdot\left(h^{\gamma}\right)^{\mathcal{E}_{n-1}}
\right)^k
\end{equation}

\begin{align*}
\mathcal{E}_1 &= \textstyle\sum_{u \in S} \mathcal{H}\left(u\right) \\
\mathcal{E}_2 &= \textstyle\sum_{u_1, u_2 \in S, u1 \neq u2} \mathcal{H}\left(u_1\right) \cdot \mathcal{H}\left(u_2\right) \\
&... \\[0in]
\mathcal{E}_{n-1} &= \textstyle\prod_{u \in S}\mathcal{H}(u)
\end{align*}

For IBBE, computing $C_2$ is bound by the computations of all $\mathcal{E}$, thus requires a quadratic number of operations $O\left(|S|^2\right)$.
In the case of IBBE-SGX, having access to $M_{SK}$ allows computing $C_2$ directly using Formula~\ref{c_2}.
It thus requires a linear number of operations.

Moreover, we augment the ciphertext values with $C_3$, which will prove useful for the subsequent operations:

\begin{equation}
C_3 = h^{\prod\limits_{u \in S}\left(\gamma + \mathcal{H}\left(u\right)\right)}
\end{equation}

Note that $C_3$ can be stored publicly as it can be computed entirely from $PK$.
 
\subsection{Decrypt Broadcast Key}\label{sec:appendix:decrypt}
The decrypt operation is executed identically for IBBE and IBBE-SGX, and relies on $PK$. 
A user can make use of her secret key $u_{SK}$ to compute $b_k$, given $\left(\mathcal{S}, \mathcal{C}\right)$
We chose to omit presenting the intricate formula as we maintain it in the original form, as shown in ~\cite{Delerablee:2007:IBBE}.

\subsection{Add User to Broadcast Key}\label{sec:appendix:add}
As the joining user $u_{add}$ is allowed to decrypt group secrets prior to joining, there is no need of a re-key operation by changing the value of $b_k$.
The only required change is therefore to incorporate $u_{add}$ into  $\mathcal{S}$, and $\mathcal{H}\left(u_{add}\right)$ into $C_2$.

For IBBE, including $\mathcal{H}(u_{add})$ into all $\mathcal{E}$ values requires a quadratic number of operations. 
For IBBE-SGX, by making use of $M_{SK}$, one has access to $\gamma$, thus the new user is included in constant time:
$ C_2 \gets \left(C_2\right)^{\gamma + \mathcal{H}\left(u_{add}\right)} $.

\subsection{Remove User form Broadcast Key}\label{sec:appendix:remove}

Whenever removing a user $u_{rem}$, all group elements  $b_k, \mathcal{S}$ and $\mathcal{C}$ need to change.  
$b_k$ and $C_1$ can be computed by Formulas \ref{K_from_k} and \ref{c_1}, once a new random value for $k \in \mathbb{Z}_p^*$ is picked.

Within the traditional assumption, $C_2$ is computed similarly to encrypting group key operation, consuming a quadratic number of operations. 
Within IBBE-SGX, having access to $\gamma$ through the $M_{SK}$ allows first changing $C_3$ and then $C_2$ in constant time:

\begin{align}
C_3 &\gets \left(C_3\right)^{\left(\gamma + \mathcal{H}\left(u_{rem}\right)\right)^{-1}} \\
C_2 &\gets \left(C_3\right)^k \label{C2_gets_C3}
\end{align}

\subsection{Re-key Broadcast Key}\label{sec:appendix:rekey}

Sometimes, it is necessary to change the value of $b_k$ without performing any group membership changes.
This re-keying operation can be performed optimally in constant time under both usage model assumptions, by making use of $C_3$. 

First, a new random $k \in \mathbb{Z}_p^*$ is generated and the new key computed by Formula~\ref{K_from_k}. 
$C_1$ can be computed by Formula~\ref{c_1}, while $C_2$ is computed from $C_3$ by~Formula \ref{C2_gets_C3}.

\end{document}